\documentclass[sigconf]{acmart}
\AtBeginDocument{%
  \providecommand\BibTeX{{%
    \normalfont B\kern-0.5em{\scshape i\kern-0.25em b}\kern-0.8em\TeX}}}

\copyrightyear{2023}
\acmYear{2023}
\setcopyright{rightsretained}
\acmConference[CIKM '23]{Proceedings of the 32nd ACM International Conference on Information and Knowledge Management}{October 21--25, 2023}{Birmingham, United Kingdom}
\acmBooktitle{Proceedings of the 32nd ACM International Conference on Information and Knowledge Management (CIKM '23), October 21--25, 2023, Birmingham, United Kingdom}\acmDOI{10.1145/3583780.3615004}
\acmISBN{979-8-4007-0124-5/23/10}

\usepackage{etoolbox}
\makeatletter
\patchcmd{\maketitle}{\@copyrightpermission}{
  \begin{minipage}{0.3\columnwidth}
     \href{https://creativecommons.org/licenses/by/4.0/}{\includegraphics[width=0.90\textwidth]{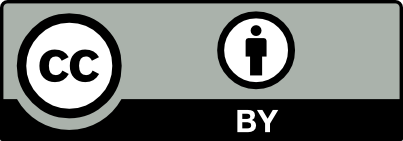}}
  \end{minipage}\hfill
  \begin{minipage}{0.7\columnwidth}
     \href{https://creativecommons.org/licenses/by/4.0/}{This work is licensed under a Creative Commons Attribution International 4.0 License.}
  \end{minipage}
  
  \vspace{5pt}
}{}{}
\makeatother
\usepackage{diagbox}

\usepackage{float}
\usepackage{array}
\usepackage{multirow}
\usepackage{threeparttable}
\usepackage{subfigure}
\usepackage{bbding}
\usepackage{enumitem}
\setlist[itemize]{leftmargin=*}
\begin{document}

%%
%% The "title" command has an optional parameter,
%% allowing the author to define a "short title" to be used in page headers.
\title{Parallel Knowledge Enhancement based Framework for Multi-behavior Recommendation}

%%
%% The "author" command and its associated commands are used to define
%% the authors and their affiliations.
%% Of note is the shared affiliation of the first two authors, and the
%% "authornote" and "authornotemark" commands
%% used to denote shared contribution to the research.

\author{Chang Meng}
\authornote{Both authors contributed equally to this research.}
\email{mengc21@mails.tsinghua.edu.cn}
\affiliation{
  \institution{Shenzhen International Graduate School, Tsinghua University}
  \city{Shenzhen}
  \country{China}
}

\author{Chenhao	Zhai}
\authornotemark[1]
\email{zhaich2216@gmail.com}
\affiliation{
  \institution{Shenzhen International Graduate School, Tsinghua University}
  \city{Shenzhen}
  \country{China}
}

\author{Yu Yang}
\email{yy286010606@gmail.com}
\affiliation{
  \institution{Shenzhen International Graduate School, Tsinghua University}
  \city{Shenzhen}
  \country{China}
}

\author{Hengyu Zhang}
\email{zhang-hy21@mails.tsinghua.edu.cn}
\affiliation{
  \institution{Shenzhen International Graduate School, Tsinghua University}
  \city{Shenzhen}
  \country{China}
}

\author{Xiu Li}
\authornote{The corresponding author.}
\email{li.xiu@sz.tsinghua.edu.cn}
\affiliation{
  \institution{Shenzhen International Graduate School, Tsinghua University}
  \city{Shenzhen}
  \country{China}
}

\begin{abstract}
Multi-behavior recommendation algorithms aim to leverage the multiplex interactions between users and items to learn users' latent preferences. Recent multi-behavior recommendation frameworks contain two steps: fusion and prediction. In the fusion step, advanced neural networks are used to model the hierarchical correlations between user behaviors. In the prediction step, multiple signals are utilized to jointly optimize the model with a multi-task learning (MTL) paradigm. However, recent approaches have not addressed the issue caused by imbalanced data distribution in the fusion step, resulting in the learned relationships being dominated by high-frequency behaviors. In the prediction step, the existing methods use a gate mechanism to directly aggregate expert information generated by coupling input, leading to negative information transfer. To tackle these issues, we propose a Parallel Knowledge Enhancement Framework (PKEF) for multi-behavior recommendation. Specifically, we enhance the hierarchical information propagation in the fusion step using parallel knowledge (PKF). Meanwhile, in the prediction step, we decouple the representations to generate expert information and introduce a projection mechanism during aggregation to eliminate gradient conflicts and alleviate negative transfer (PME). We conduct comprehensive experiments on three real-world datasets to validate the effectiveness of our model. The results further demonstrate the rationality and effectiveness of the designed PKF and PME modules. The source code and datasets are available at \url{https://github.com/MC-CV/PKEF}.
\end{abstract}

%%
%% The code below is generated by the tool at http://dl.acm.org/ccs.cfm.
%% Please copy and paste the code instead of the example below.
%%


%%
%% Keywords. The author(s) should pick words that accurately describe
%% the work being presented. Separate the keywords with commas.

%% A "teaser" image appears between the author and affiliation
%% information and the body of the document, and typically spans the
%% page.

\maketitle
\section{Introduction}
\label{intro}

Recommender systems are information filtering techniques designed to provide personalized services based on user preferences. In our daily lives, recommendation systems are widely used in various scenarios such as e-commerce, social media, music, and video platforms. Early collaborative filtering (CF) techniques \cite{cfsurvey} made recommendations based on users' historical interactions with items, but they had limitations in effectively utilizing diverse user behavior information for recommendations. In the real world, user behavior goes beyond a single type and includes various behaviors such as viewing, adding to cart, and purchasing. Among them, we mainly focus on a specific behavior, namely target behavior (e.g., \textit{buy}) and considering other behaviors as auxiliary behaviors \cite{nmtr,matn,mbgcn}. These multiple behavior signals carry rich user preferences, which can be leveraged to comprehensively understand user needs and provide better services.

Recent researches have focused on effectively leveraging multiple behavior signals for recommendations. Existing frameworks for multi-behavior recommendation contain two steps: multi-behavior fusion and multi-behavior prediction \cite{he2023survey}. In the fusion step, advanced neural networks are applied to capture the correlations between users and items across multiple behaviors. In the prediction step, multi-task learning (MTL) is devised to further utilize the heterogeneous interaction information \cite{huang2021recent}.

\textbf{Multi-behavior Fusion.} Early studies applied matrix factorization \cite{mf1,mf2,mf3} to multi-behavior recommendation. With the rise of deep learning, neural network-based approaches \cite{matn,nmtr,dipn} have become popular in multi-behavior fusion. These methods can model the complex relationships between users and items, capturing richer user interests and item features. Among them, graph neural networks \cite{lightgcn,ngcf,lr-gccf,cigf,ckml,hpmr} have been widely applied in multi-behavior recommendation due to their ability to efficiently utilize high-order connectivity between users and items \cite{ghcf,khgt}. For example, MBGCN \cite{mbgcn} and GNMR \cite{gnmr} utilize graph neural networks to improve recommendation performance. However, these methods do not consider using dependencies between behaviors to assist model learning. In the real world, user behaviors often follow a hierarchical order, such as \textit{view → cart → buy}. User preference information from upstream behaviors (e.g., \textit{view}) can be used to assist downstream tasks (e.g., \textit{cart} and \textit{buy}) \cite{nmtr,crgcn,mbcgcn}. CRGCN \cite{crgcn} and MB-CGCN \cite{mbcgcn} integrate the cascade dependencies between behaviors into graph convolutional networks (GCNs), facilitating the learning of user and item embeddings. These models, which consider behavioral hierarchy, have demonstrated better performance compared to previous approaches.

\textbf{Multi-behavior Prediction.} Multi-task learning (MTL) is a commonly used approach in multi-behavior prediction, as it can effectively utilize complex heterogeneous signals from multiple tasks to jointly optimize the model. Existing multi-task learning methods typically have coupled inputs for different tasks \cite{sharebottom,mmoe,PLE}. They generate multiple experts in different ways and aggregate the expert information for subsequent tasks. For example, MMOE \cite{mmoe} utilizes coupled representations to generate multiple experts and assigns learnable weights to each task to aggregate the expert information. PLE \cite{PLE} further improves this approach by generating the specific experts for each task on the basis of the shared experts for all tasks.

\begin{figure}[!t]
	\setlength{\belowcaptionskip}{-0cm}
	\setlength{\abovecaptionskip}{-0.1cm}
	\subfigure{
        \begin{minipage}[t]{0.3\linewidth}
        \centering
		\label{fig:distribution_beibei} 
		\includegraphics[width=1.0in]{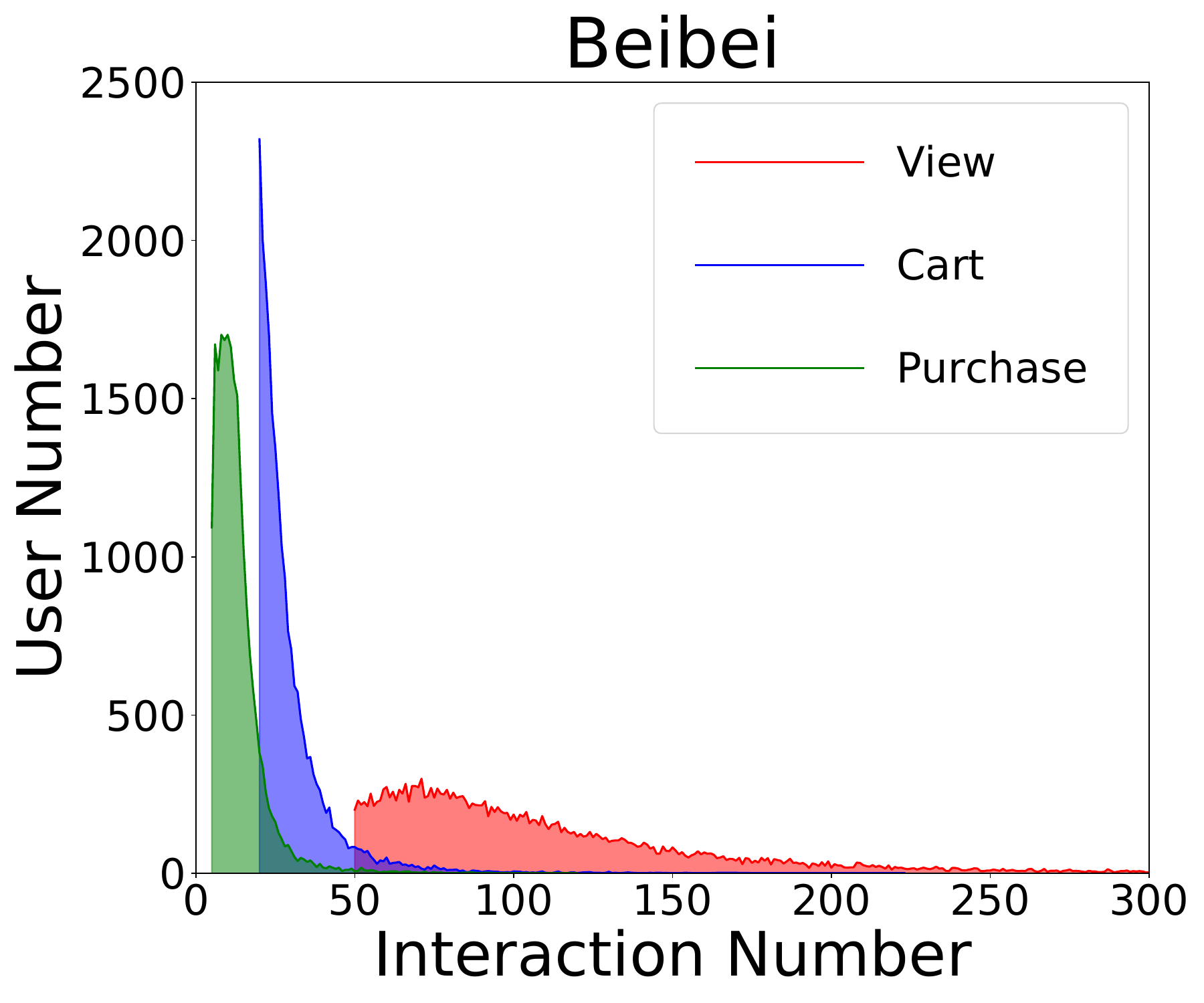}
        \end{minipage}}
	\subfigure{
        \begin{minipage}[t]{0.3\linewidth}
        \centering
		\label{fig:distribution_taobao} 
		\includegraphics[width=1.0in]{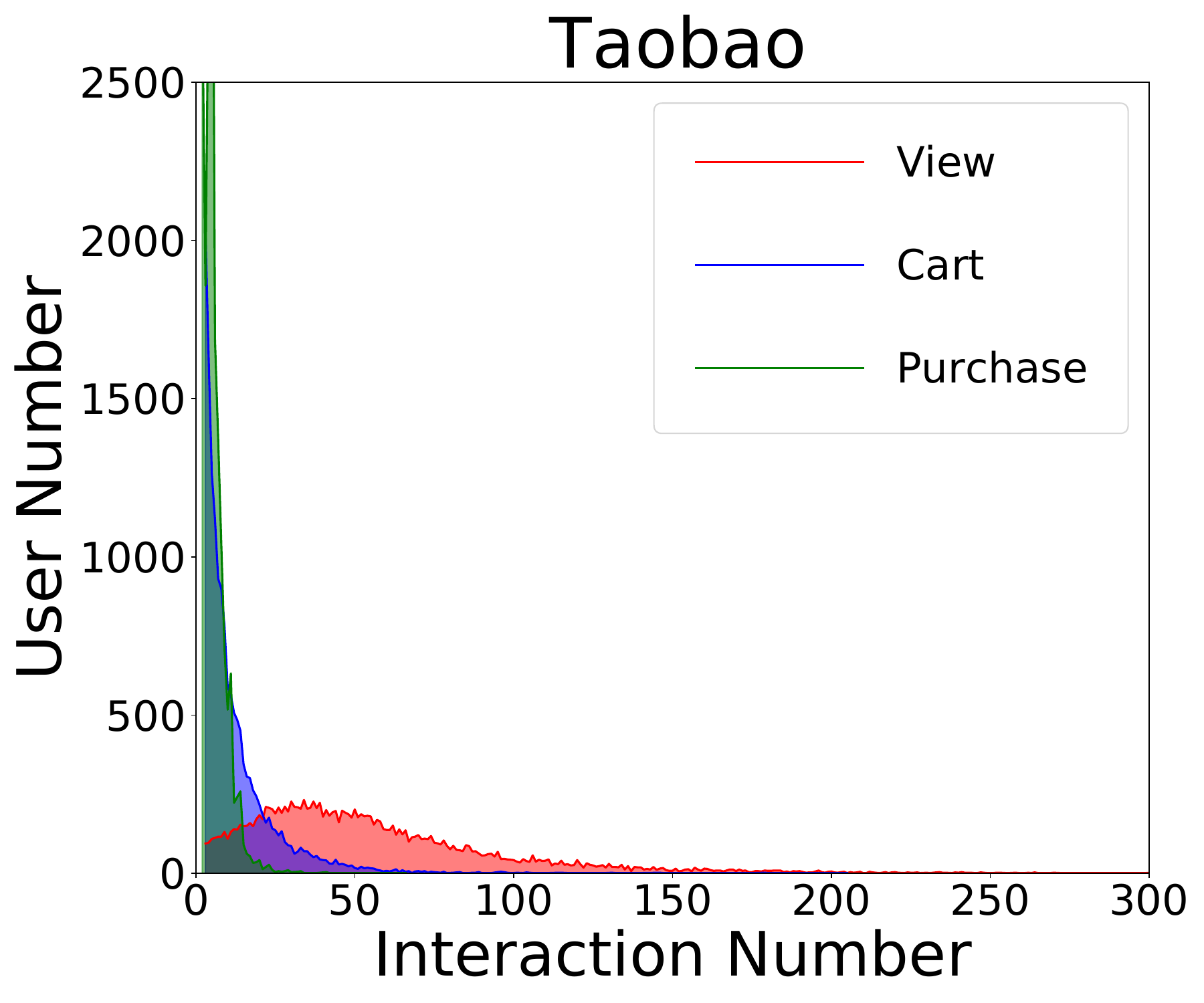}
        \end{minipage}}
    \subfigure{
        \begin{minipage}[t]{0.3\linewidth}
        \centering
		\label{fig:distribution_tmall} 
		\includegraphics[width=1.0in]{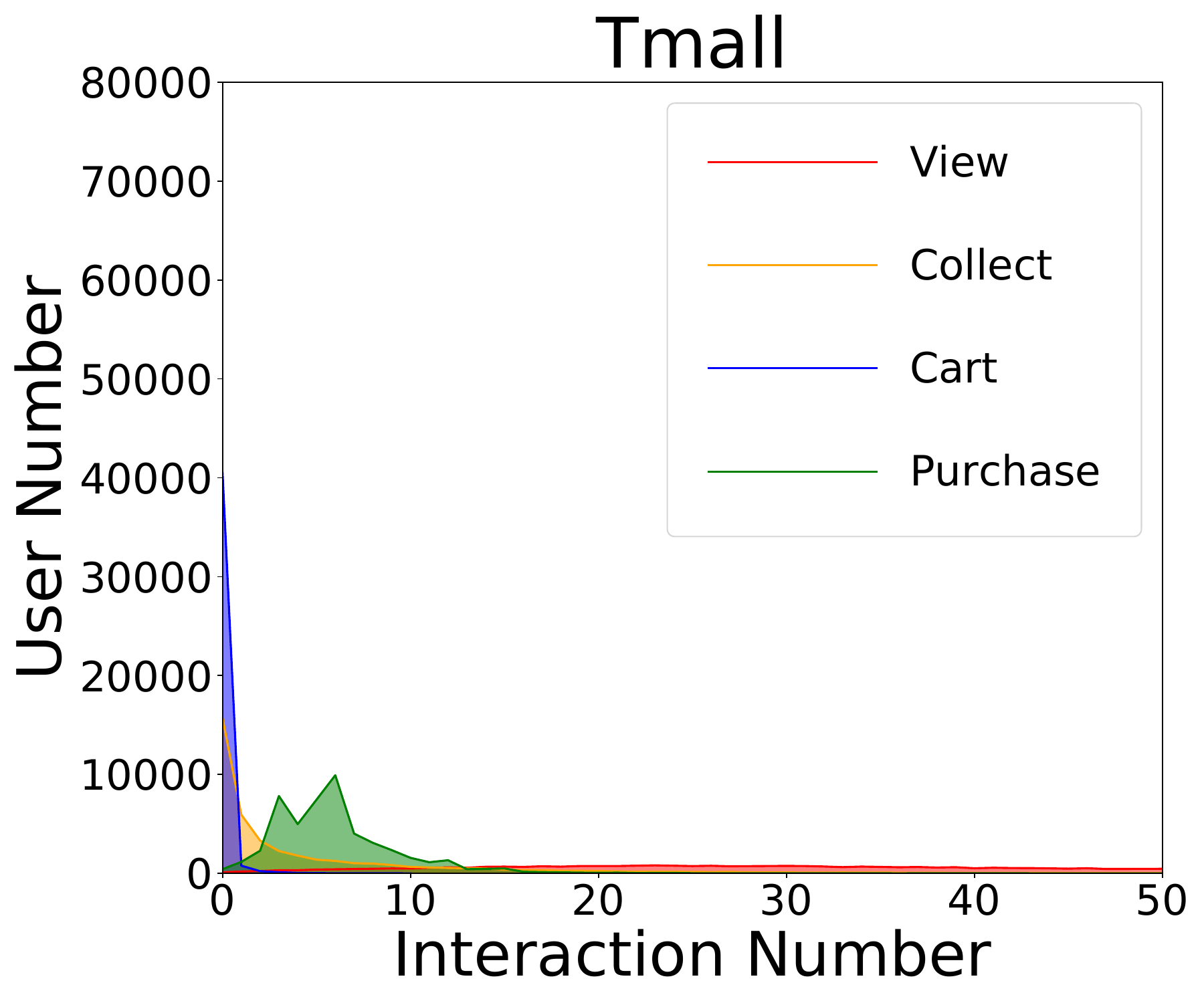}
        \end{minipage}}
	\caption{Histogram of user numbers w.r.t interaction numbers for different behaviors.}
	\vspace{-3mm}
	\label{fig:distribution}
\end{figure}

\begin{figure}[t]
	\centering
	\setlength{\belowcaptionskip}{0cm}
	\setlength{\abovecaptionskip}{0cm}
	\includegraphics[width=0.48\textwidth]{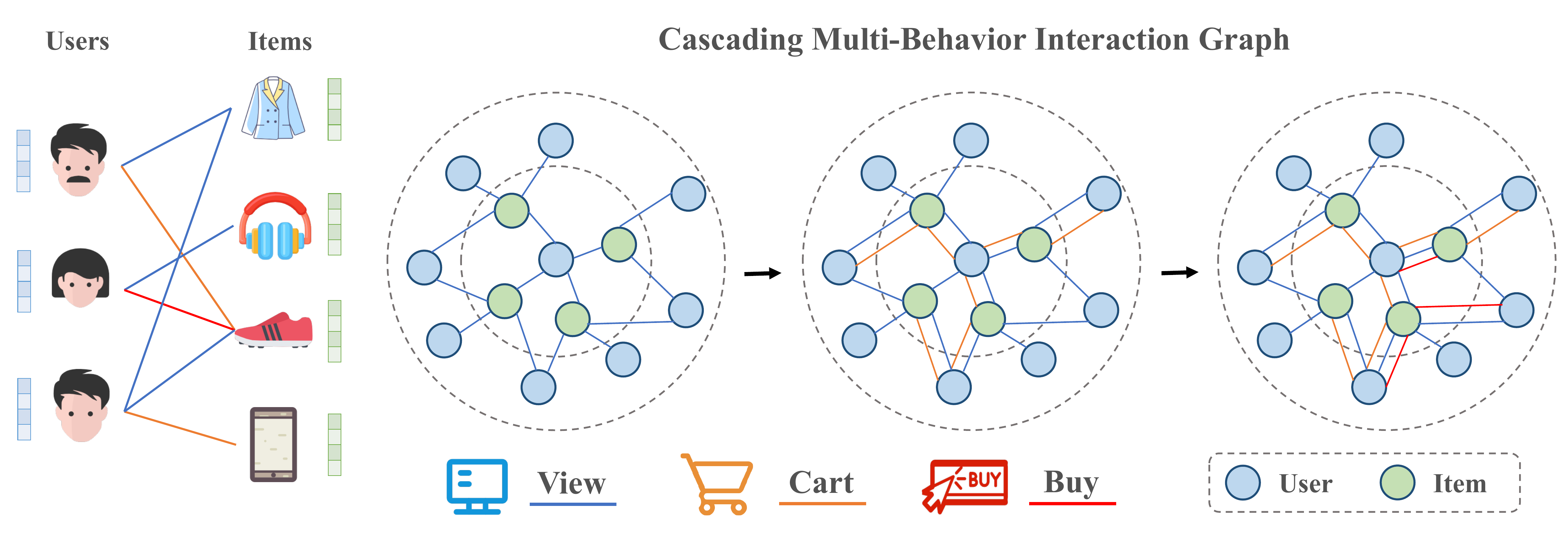}
	\caption{Illustration of the learned multi-behavioral correlations in the cascade stream.}
	\label{fig:cascading_relation}
	\vspace{-3mm}
\end{figure}

Although cascade graph convolutional networks and MTL-based multi-behavior recommendation methods have made significant progress for multi-behavior fusion and prediction respectively, they also have their limitations:
\begin{itemize}
\item \textbf{Ignorance of the imbalanced behavioral distribution.} As shown in Figure \ref{fig:distribution} (plotting the data distribution), the interactions for different behaviors are highly imbalanced. One behavior (e.g., \textit{view}) may account for the majority of the total interactions. In the cascade behavior modeling, this imbalance problem is further exacerbated. 
As shown in Figure \ref{fig:cascading_relation}, in the cascade stream, upstream behaviors have richer interaction information compared to downstream behaviors. Thus, in the process of behavior propagation, the learned relationships are dominated by upstream behaviors, leading to a biased relationship learned by the model towards upstream behaviors, which interferes with downstream behavior prediction.
% Figure \ref{} illustrates this for one user, showing high-order information interaction graphs at different levels (\textit{view → cart → buy}). It can be observed that upstream behaviors have richer interaction information compared to downstream behaviors. Thus, in the process of behavior propagation, the learned interaction information is dominated by upstream behaviors, leading to a biased relationship learned by the model towards upstream behaviors, which interferes with downstream behavior prediction.

\item \textbf{Negative transfer problem.} When training multiple tasks, the performance of certain tasks can be negatively affected or interfered with by other tasks, resulting in performance degradation. This is known as the negative transfer phenomenon \cite{negativeTransfer}. In multi-task learning, although \textit{coupled inputs} can share information from different behaviors, they can also introduce potential gradient conflict issues (explained in Section \ref{problem_gradient}). Additionally, when aggregating expert information from different behaviors for a specific task, noise from other behaviors is often introduced, leading to negative transfer problems.
\end{itemize}

To address these two issues, we propose a \textbf{\underline{P}}arallel \textbf{\underline{K}}nowledge \textbf{\underline{E}}nhancement based \textbf{\underline{F}}ramework (PKEF) for
Multi-behavior Recommendation. It consists of the Parallel Knowledge Fusion module (PKF) and the Projection Disentangling Multi-Experts
network (PME). To address the first issue, PKF combines the cascade and parallel paradigms, leveraging parallel knowledge for adaptive enhancement of different behaviors' representations while learning hierarchical correlation information to correct the information bias caused by imbalanced behavioral interactions.

To address the second issue, 
% the PME module resolves the gradient conflict introduced by coupled inputs in neural architectures similar to MMOE \cite{mmoe} and PLE \cite{PLE}. 
PME regards different behaviors as independent tasks, generates corresponding expert informations for each behavior with \textit{separate inputs}, and aggregates the expert information from different behaviors using learnable weights. Considering that the aggregation of different behaviors may introduce noise during the learning process for a specific behavior task, PME introduces a projection mechanism during aggregation to disentangle the shared and unique parts for other behavioral experts. The shared part is used for aggregation, avoiding the introduction of harmful information. For the unique part, an auxiliary loss is designed for optimizing, which promotes the effectiveness of complementary shared information. PME alleviates the negative transfer phenomenon while solving the gradient conflict problem (explained in Section \ref{PME_gradient}).

In summary, our work makes the following contributions:
\begin{itemize}
    \item We investigate the issues of ignorance of the imbalanced behavioral distribution in the cascade paradigm of multi-behavior recommendation and the negative transfer phenomenon in MTL. We propose an innovative multi-behavior recommendation framework (PKEF) to address these issues. It consists of the Parallel Knowledge Fusion module (PKF) and the Projection Disentangling Multi-Expert network (PME).
    \item To achieve better recommendation performance, we address the issue of imbalanced data distribution for different behaviors by enhancing the hierarchical information propagation in the cascade process using parallel knowledge (PKF). Additionally, we alleviate the gradient conflict introduced by coupled MTL inputs and propose a projection-based denoising method to remove harmful information between behaviors, effectively solving the negative transfer problem (PME).
    \item We conduct comprehensive experiments on three real-world datasets to demonstrate the effectiveness of our model. Further experimental results verify the rationality and effectiveness of the designed PKF and PME modules.
\end{itemize}

\section{Related Work}
\label{related_work}

\textbf{Multi-behavior Recommendation.}
Multi-behavior recommendation methods use multiple user-item interactions to solve the data sparsity problem. In recent years, this approach has attracted widespread attention.

Early multi-behavior recommendation methods usually handle multi-behavior data by introducing multiple matrix factorization \cite{mf1,mf2,mf3} or designing new sampling strategies \cite{bprh,mc-bpr,bpr_resolve}. The former one extends the traditional matrix factorization technique by conducting it on multiple matrices with shared embeddings, such as CMF \cite{cmf}. The latter one uses multiple behaviors as auxiliary data and designs new sampling strategies to enrich the training samples, such as MF-BPR \cite{mc-bpr} and VALS \cite{vals}, which introduce and improve negative sampling strategies.

With the development of deep learning techniques \cite{chengqing2023multi,zhou2023artificial,liang2023knowledge}, researchers have started to explore multi-behavior recommendation models based on deep neural network (DNN) or graph convolutional network (GCN). DNN-based models usually design models to learn embeddings from each behavior and integrating them into the prediction of target behaviors. For example, DIPN \cite{dipn} and MATN \cite{matn} use different attention mechanisms to model the relationship between behaviors for embedding learning and aggregation. NMTR \cite{nmtr} differs from the above methods by using a multi-task learning model in which all behaviors of the users serve as prediction targets and the prediction scores of the previous behavior are passed to the next behavior for prediction.

GCN-based models learn user embeddings by constructing a unified user-item graph and performing graph convolution operations. GHCF \cite{ghcf} explicitly models the high-order relationship between users and items through GCN and performs multi-task learning to predict each behavior through a non-sampling approach. MBGCN \cite{mbgcn} takes behavior semantic into account, capturing it by item-item propagation layer and combines behavior semantic with behavior contributions learned from user-item propagation layer for score prediction. The recently proposed CRGCN \cite{crgcn} and MB-CGCN \cite{mbcgcn} take into account the hierarchical correlation between behaviors and achieve great performance by building cascaded graph convolutional networks to capture user preferences. However, due to the imbalanced distribution of the interactions among different behaviors, simply employing cascaded networks will lead to the learned relationships being dominated by high-frequency behaviors, which interferes with downstream behavior prediction.
% can lead to the transmission of noisy behaviors in upstream behaviors, which affects the recommendation effect.

\textbf{MTL for Recommendation.}
With the growing diversity of user interests, the limitations of single-task learning in traditional recommender systems have become more and more obvious, especially in the face of multiple signals. To solve the above dilemma, in recent years, researchers have attempted to apply multi-task learning to recommender systems. One model widely used in multi-behavior recommendation is the shared bottom \cite{sharebottom} structure, where each task shares the same bottom parameters to extract common features, while the parameters at the top layer are independent. However, approaches based on this structure \cite{ghcf,nmtr,mbgmn} will lead to negative transfer phenomenon and trigger a seesaw effect for tasks with weak relevance. To solve these problems, MTL structures based on gated expert algorithm are proposed. MOE \cite{moe} divides the shared bottom structure into multiple experts that learn different features separately. MMOE \cite{mmoe} extends MOE by introducing a task-specific gating mechanism to obtain different fusion weights in multi-task learning. PLE \cite{PLE} further proposes to employ shared or task-specific experts at the bottom layer and combine them adaptively through gating networks. However, these methods use coupled inputs for multiple tasks, which leads to the gradient conflict problem and negative transfer phenomenon, thus affecting the model performance (Illustrated in Section \ref{problem_gradient}). 

\section{Problem Definition}

\subsection{Problem Definition}
\label{Definition}
We define $u$ and $v$ as a user and an item, respectively. Meanwhile, $\mathbf{U}$ and $\mathbf{V}$ denote the user and item sets, respectively. The adjacency matrices of multiple behaviors can be represented by a set, i.e., $\mathcal{M} =\left\{\mathbf{M}_{1},\mathbf{M}_{2},\cdots,\mathbf{M}_{K}\right\}$, where $\mathbf{M}_{k}=\left[m_{(k)uv}\right]_{|\mathbf{U}|\times|\mathbf{V}|}\in \left\{0, 1\right\}$ indicates whether the user $u$ interacted with the item $v$ under behavior $k$. Furthermore, in order to represent the heterogeneous interaction information of users and items more conveniently, we define the multiplex user-item bipartite graph $\mathcal{G}=(\mathcal{H}, \mathcal{E}, \mathcal{M})$, where $\mathcal{H} = \mathbf{U}\cup\mathbf{V}$, $\mathcal{E} = \cup_{k = 1}^{K}\mathcal{E}_{k}$ is the edge set including all behavior records between users and items. In the multi-behavior recommendation, we assume that $k \in \{1,2,...,K\}$, and the number corresponds to the upstream and downstream relationships between behaviors. The larger the number, the more downstream the behavior (i.e., $K$ is the most downstream behavior). Last but not least, there exists a target behavior (denotes as $\mathbf{M}_{K}$) to be optimized, which is purchasing (buying) for e-commerce scenarios.

\subsection{Gradient Issue in MTL}

\subsubsection{Gradient Conflict with Coupled Input}
\label{problem_gradient}
Most of the existing methods use the coupled input for MTL, as summarized in Section \ref{related_work}. 
This may cause a gradient conflict issue in MTL which restricts their learning ability for each task. As the classical MTL methods directly couple the representations of different behaviors together with different weights, we have: 
\begin{equation}
\setlength{\abovedisplayskip}{0.5pt}
\setlength{\belowdisplayskip}{0.5pt}
\begin{aligned}
\mathbf{e}_{u}^{*} = \sum_{k=1}^{K}\lambda_{k}\mathbf{e}_{u}^{k}, \mathbf{e}_{v}^{*} = \sum_{k=1}^{K}\lambda_{k}\mathbf{e}_{v}^{k},
\end{aligned}
\end{equation}
% $\mathbf{e}_{u}^{*}$ and $\mathbf{e}_{i}^{*}$ denote the learned embedding of user $u$ and item $i$, and
where $K$ is the number of behaviors, $\lambda_{k}$ is the weight of $k$-th behavior.
% embedding which contains all behaviors’ input information, respectively. 
Taking $(\mathbf{e}_{u}^{*},\mathbf{e}_{v}^{*})$ as input for MTL, the loss function can be formulated as:
% As we can see, in the existing MTL models, all the behavior share the same representation vector. Then we consider the loss function, which can be formulated as:
\begin{equation}
\setlength{\abovedisplayskip}{0.5pt}
\setlength{\belowdisplayskip}{0.5pt}
\begin{aligned}
\mathcal{L}_{uv}
% &=\sum_{k=1}^{K}L(\hat{o}_{uv}^{k}-{o}_{uv}^{k})\\
&=\sum_{k=1}^{K}L(f_{k}(\mathbf{e}_{u}^{*},\mathbf{e}_{v}^{*})-{o}_{uv}^{k}),
\end{aligned}
\end{equation}
where $\hat{o}_{uv}^{k}$ denotes the predictive probability that user $u$ will interact with item $v$ under the \textsl{k}-th behavior, ${o}_{uv}^{k}$ is the true label, $L(\cdot)$ is the loss function, and $f_{k}(\cdot)$ is the predictive function in MTL models.
Then we have:
\begin{equation}
\setlength{\abovedisplayskip}{0.5pt}
\setlength{\belowdisplayskip}{0.5pt}
% \begin{aligned}
{\partial{\mathcal{L}_{uv}}\over{\partial{(\mathbf{e}_{u}^{*} \circ \mathbf{e}_{v}^{*})}}}
% &=
% \sum_{k=1}^{K}{\partial{L(f_{k}(\mathbf{e}_{u}^{*},\mathbf{e}_{v}^{*})-{o}_{uv}^{k})}
% \over{\partial{(\mathbf{e}_{u}^{*} \circ \mathbf{e}_{v}^{*})}}}\\
=
\sum_{k=1}^{K}{\partial{f_{k}(\mathbf{e}_{u}^{*},\mathbf{e}_{v}^{*})}\over{\partial{(\mathbf{e}_{u}^{*} \circ \mathbf{e}_{v}^{*})}}}*{L^{'}(f_{k}(\mathbf{e}_{u}^{*},\mathbf{e}_{v}^{*})-{o}_{uv}^{k})}
=
\sum_{k=1}^{K}{a_{uv}^{k}{\mathbf{r}^{k}}},
% &=
% \sum_{k=1}^{K}{\mathbf{r}^{' k}}
% \end{aligned}
\end{equation}
where ($\circ$) is the hadamard product operation, $a_{uv}^{k}=L^{'}(f_{k}(\mathbf{e}_{u}^{*},\mathbf{e}_{v}^{*})-{o}_{uv}^{k})$ is a scalar. $\mathbf{r}^{k}={\partial{f_{k}(\mathbf{e}_{u}^{*},\mathbf{e}_{v}^{*})}\over{\partial{(\mathbf{e}_{u}^{*} \circ \mathbf{e}_{v}^{*})}}}$.
As $\mathbf{r}^{k}$ denotes the derivative of a scalar to a vector, it is also a vector.
$\forall$ $k \in \{1,2, \ldots, K\}$, $a_{uv}^{k}{\mathbf{r}^{k}}$ determines the updating magnitude and direction of the vector $\mathbf{e}_{u}^{*} \circ \mathbf{e}_{v}^{*}$.
We can see that the gradients from all behaviors are coupled, and they jointly optimize the same vector $\mathbf{e}_{u}^{*} \circ \mathbf{e}_{v}^{*}$, which leading to gradient conflicts. As a result, the harmful information coupled in the input affects the learning of the target behavior information in the training process, leading to negative transfer.
% performance deterioration when information is transferred across different tasks through the classical MTL structure.

\subsubsection{Projection Disentangling Multi-Experts with Separated Input}
\label{PME_gradient}
To handle the above problem, we first need to utilize the separated input of each behavior to generate the behavior-specific expert information and behavior-specific gating weight. Thus, we have:
\begin{equation}
\mathbf{e}_{uv}^{k} = \mathbf{e}_{u}^{k} \circ \mathbf{e}_{v}^{k}, \hat{\mathbf{g}}_{exp}^{k} = FC_g^{k}(\mathbf{e}_{u}^{k} || \mathbf{e}_{v}^{k})
\end{equation}
where $\hat{\mathbf{g}}_{exp}^{k} \in \mathbb{R}^{K \times 1}$ is the weight of gate. $FC_g^{k}(\cdot)$ represents the behavior-specific fully connected layer.
Further, while aggregating the information from different experts, the gating mechanism simultaneously introduce the negative information from other experts. Thus, we need to extract the information that is useful to the prediction of behavior $k$ from other experts ($\mathbf{e}_{uv}^{k^{\prime}}$). In details, we leverage a projection mechanism and have:
\begin{equation}
\setlength{\abovedisplayskip}{0.5pt}
\setlength{\belowdisplayskip}{0.5pt}
\mathbf{e}_{sha}^{k^{\prime},k}=a_{k^{\prime},k}\mathbf{e}_{uv}^{k}, 
\hat{\mathbf{e}}_{exp}^k=\mathop{Concat}\limits_{k^{\prime}=1}^K(\mathbf{e}_{sha}^{k^{\prime},k})
\end{equation}
where $\mathbf{e}_{sha}^{k^{\prime},k}$ represents the shared information extracted from $\mathbf{e}_{uv}^{k^{\prime}}$ with the guidance of $\mathbf{e}_{uv}^{k}$. $a_{k^{\prime},k} = \frac{\mathbf{e}_{uv}^{k^{\prime}}\cdot\mathbf{e}_{uv}^{k}}{|\mathbf{e}_{uv}^{k}||\mathbf{e}_{uv}^{k}|}$ is a scalar and can be flexibly adjusted. 
In practice, we adjust the scalar by multiplying by a small value.

Finally, we analyse the optimization of input $(\mathbf{e}_{u}^{t} \circ \mathbf{e}_{v}^{t})$, and have:
\begin{equation}
\label{decoupled_input}
\begin{aligned}
{\partial{\mathcal{L}_{uv}}\over{\partial{(\mathbf{e}_{u}^{t} \circ \mathbf{e}_{v}^{t})}}}
&=\sum_{k=1}^{K} \frac{\partial \hat{f}_{k}(\hat{\mathbf{g}}_{exp}^{k}, \hat{\mathbf{e}}_{exp}^k)}{\partial(\mathbf{e}_{u}^{t} \circ \mathbf{e}_{v}^{t})}\\
&=\sum_{k=1}^{K} \frac{\partial \hat{f}_{k}(\hat{\mathbf{g}}_{exp}^{k}, \mathbf{e}_{uv}^k)}{\partial(\mathbf{e}_{u}^{t} \circ \mathbf{e}_{v}^{t})}
=\frac{\partial \hat{f}_{t}(\hat{\mathbf{g}}_{exp}^{t}, \mathbf{e}_{uv}^t)}{\partial(\mathbf{e}_{u}^{t} \circ \mathbf{e}_{v}^{t})}
\end{aligned}
\end{equation}
where $\hat{f}_{k}(a,b)$ means the expressions with respect to variables $a$ and $b$ under behavior $k$. Without loss of generality, we can find that $\forall t \in \{1,2,\cdots,K\}$, the gradient of each behavior optimizes along the direction of their respective input (e.g., the gradient of the $t$-th behavior optimizes $\mathbf{e}_{u}^{t} \circ \mathbf{e}_{v}^{t}$ independently), so that the gradient conflicts problem is successfully solved.
\section{METHODOLOGY}
\begin{figure*}[t]
	\centering
	\setlength{\belowcaptionskip}{-0.0cm}
	\setlength{\abovecaptionskip}{-0.0cm}
	\includegraphics[width=0.88\textwidth]{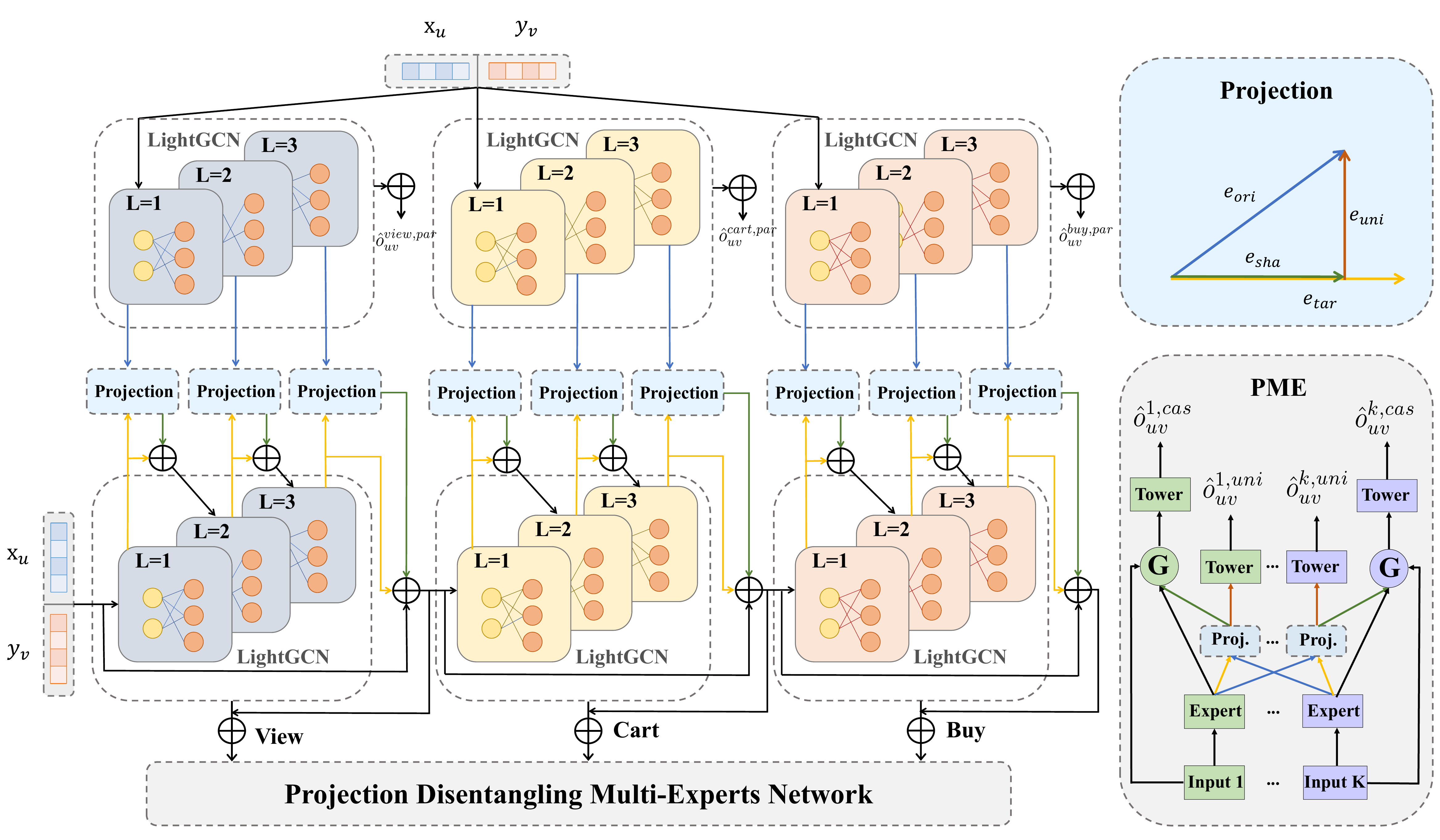}
	\caption{Illustration of the proposed PKEF framework. ($\oplus$) denotes the element-wise addition operation. Lines of different colors correspond to representations of different colors in the Projection module (e.g., blue lines denote $e_{ori}$ and green lines represent $e_{sha}$). For brevity, we illustrate the knowledge fusion between cascade and parallel stream with a projection scheme.}
	\label{fig:framework}
	\vspace{-3mm}
\end{figure*}

We devise a "Parallel Knowledge Enhancement based Framework" (PKEF) for multi-behavior recommendation, which contains two parts: (1) Parallel Knowledge Fusion (PKF) module; (2) Projection Disentangling Multi-Experts (PME) network. Figure \ref{fig:framework} illustrates the technical details of the proposed framework.

\subsection{Embedding Layer}
In industrial applications, users and items are often denoted as high-dimensional one-hot vectors. However, to transform the high-dimensional sparse vectors into low-dimensional dense embeddings, we apply the embedding lookup operation for user $u$ and item $v$ to obtain the embedding vectors. Generally, we have:
\begin{equation}
\begin{aligned}
\mathbf{E}_u=[\underbrace{\mathbf{x}_{u_1}, \cdots, \mathbf{x}_{u_{|\mathbf{U}|}}}_{\text {users embeddings }}],&
\mathbf{x}_u=\operatorname{LookUp}\left(u, \mathbf{E}_u\right)\\
\mathbf{E}_v=[\underbrace{\mathbf{y}_{v_1}, \cdots, \mathbf{y}_{v_{|\mathbf{V}|}}}_{\text {item embeddings }}],&
\mathbf{y}_v=\operatorname{LookUp}\left(v, \mathbf{E}_v\right)
\end{aligned}
\end{equation}
where $\mathbf{E}_u \in \mathbb{R}^{|\mathbf{U}| \times d}$ and $\mathbf{E}_v \in \mathbb{R}^{|\mathbf{V}| \times d}$ are the embedding tables for users and items, respectively, $|\mathbf{U}|$ and $|\mathbf{V}|$ are the total number of users and items. $\mathbf{x}_u \in \mathbb{R}^{d}$ and $\mathbf{y}_v \in \mathbb{R}^{d}$ denotes the embedding vectors of user $u$ and item $v$, and $d$ is the embedding size. 

\subsection{Parallel Knowledge Fusion}
The recent multi-behavior methods ignore the imbalanced distribution of different behaviors in the fusion step. Thus the learning of these models will be more inclined to high-frequency behaviors, resulting in poor prediction effects on target behaviors.

To solve the above problem, in our model, we combine both cascade and parallel paradigms to learn complex interactions between multiple behaviors. Our Parallel Knowledge Fusion (PKF) module utilizes parallel knowledge to enhance the representations of different behaviors while learning hierarchical correlation information, so as to correct the information bias caused by the imbalance of behavior interaction distribution.
\subsubsection{Cascade Correlation Learning.}
As we have the adjacency matrices $\mathbf{M}_{1},\mathbf{M}_{2},...,\mathbf{M}_{K}$ for different behaviors, for convenience, we further process the matrices, and it can be formulated as:
\begin{equation}
\mathbf{A}_{k}=\left(\begin{array}{cc}
0 & \mathbf{M}_{k} \\
\left(\mathbf{M}_{k}\right)^{T} & 0
\end{array}\right)
\end{equation}
where $\mathbf{M}_{k}$ is the user-item adjacency interaction matrix of behavior $k$, $\mathbf{M}_{k} \in \mathbb{R}^{(|\mathbf{U}|+|\mathbf{V}|)\times(|\mathbf{U}|+|\mathbf{V}|)}$, $|\mathbf{U}|$ and $|\mathbf{V}|$ denote the number of users and items, respectively.
As graph neural networks \cite{lightgcn,ngcf} have been widely used to model the high-order interactions between users and items, we conduct a GNN-based paradigm to encode the information of each behavior. Specifically, in each behavior $k$, we apply the message passing to capture the high-order interaction information. Here, we simply leverage LightGCN \cite{lightgcn} as the GCN aggregator to aggregate information on each layer $l$:
\begin{equation}
\label{eq:mess_pass}
    \mathbf{z}^{k,l+1} = \mathbf{\hat{A}}_k\mathbf{z}^{k,l} + \mathbf{z}^{k,l}
\end{equation}
where $\mathbf{z}^{k,l}=\mathbf{z}_u^{k,l}||\mathbf{z}_v^{k,l}$. $(||)$ is the concatenate operation. $\mathbf{\hat{A}}_{k}=\mathbf{D}^{-1}(\mathbf{A}_{k}+\mathbf{I})$ is the left normalized adjacency matrix with added self-connections and $\mathbf{D}$ is a diagonal degree matrix, which is defined as $\mathbf{D}_{ii}=\sum_{j}(\mathbf{A}_{k}+\mathbf{I})_{ij}$. $\mathbf{I}$ denotes an identity matrix. And the initial $\mathbf{z}^{1,0}=\mathbf{x}_u||\mathbf{y}_v$.

Further, following MB-CGCN \cite{mbcgcn}, we conduct a cascade paradigm to learn the hierarchical correlation information of different behaviors. We have:
\begin{equation}
\mathbf{z}^{k+1,0} = \mathbf{z}^{k,L_{k}} + \mathbf{z}^{k,0}
\end{equation}
where $L_{k}$ denotes the total layers of GNN of the $k$-th behavior. Here, we apply a residual connection to combine the first and the last layer of the upstream behavior representation as the input of the downstream behavior.

\subsubsection{Parallel Interaction Enhancing.}
\label{para_inter_enhance}
In the previous part, we have modeled the hierarchical correlations of different behaviors. However, as we have illustrated in the introduction, imbalanced distribution of multiplex interactions will impact the learning of target behavior. In order to handle this problem, we further conduct a parallel learning paradigm which independently learns the representation of each behavior, and then fusioning the knowledge on each layer corresponding to the cascade stream.

Similar to the process of Equation \ref{eq:mess_pass}, we first apply the same way to each behavior, and have:
\begin{equation}
\setlength{\abovedisplayskip}{0.5pt}
\setlength{\belowdisplayskip}{0.5pt}
\label{eq:mess_pass_parallel}
    \mathbf{p}^{k,l+1} = \mathbf{\hat{A}}_k\mathbf{p}^{k,l} + \mathbf{p}^{k,l}
\end{equation}
where $\mathbf{p}^{k,l}=\mathbf{p}_u^{k,l}||\mathbf{p}_v^{k,l}$. $(||)$ is the concatenate operation. $\mathbf{\hat{A}}_{k}$ is the same as in Equation \ref{eq:mess_pass}. And the initial $\mathbf{p}^{1,0}=\mathbf{x}_u||\mathbf{y}_v$.

Then we devise two schemes to fuse the knowledge between the parallel and cascade streams, improving the Equation \ref{eq:mess_pass}. Besides, we conduct comparison experiments with other schemes (shown in Section \ref{knowledge_fusion}). 
For simplicity, we denote $\mathbf{e}_{par}^{k,l} = \mathbf{\hat{A}}_k\mathbf{p}^{k,l}$ and $\mathbf{e}_{cas}^{k,l} = \mathbf{\hat{A}}_k\mathbf{z}^{k,l}$.

\textbf{(1)} Projection-enhanced Knowledge Fusion. This scheme is inspired by DUMN \cite{dumn}, in which they used the representation projection mechanism to decouple the implicit feedback representation by the explicit feedback representation. Here, on each layer, we project the parallel representation onto the cascade representation and use the part that is collinear with it to enhance the cascade representation. It can be formulated as:
\begin{equation}
\setlength{\abovedisplayskip}{0pt}
\setlength{\belowdisplayskip}{0pt}
\left\{\begin{array}{c}
\begin{aligned}
\mathbf{p}_{col}^{k,l}&=\frac{\mathbf{e}_{par}^{k,l}\cdot\mathbf{e}_{cas}^{k,l}}{|\mathbf{e}_{cas}^{k,l}|} \frac{\mathbf{e}_{cas}^{k,l}}{|\mathbf{e}_{cas}^{k,l}|}\\
\mathbf{z}^{k,l+1}&=\mathbf{e}_{cas}^{k,l}+\mathbf{z}^{k,l}+\mathbf{p}_{col}^{k,l}
\end{aligned}
\end{array}\right.
\end{equation}
where ($\cdot$) is the vector inner product operation. $\mathbf{e}_{par}^{k,l}$ and $\mathbf{e}_{cas}^{k,l}$ are the representations of the parallel and cascade streams, respectively. $\mathbf{p}_{col}^{k,l}$ contains a mixture of behavior-specific and hierarchical correlation information.

\textbf{(2)} Vanilla-enhanced Knowledge Fusion.
Meanwhile, inspired by the vanilla attention \cite{atrank}, we devise a fusion scheme that has the similar form with it, and have:
\begin{equation}
\setlength{\abovedisplayskip}{0.5pt}
\setlength{\belowdisplayskip}{0.5pt}
\left\{\begin{array}{c}
\begin{aligned}
\mathbf{w}^{k,l} &= Softmax(\mathbf{W}_{k,l}\mathbf{e}_{cas}^{k,l} + \mathbf{b}_{k,l})\\
\mathbf{f}^{k,l}&=Concat([\mathbf{e}_{cas}^{k,l},\mathbf{e}_{par}^{k,l},\mathbf{e}_{cas}^{k,l}-\mathbf{e}_{par}^{k,l},\mathbf{e}_{cas}^{k,l} \circ \mathbf{e}_{par}^{k,l}])\\
\mathbf{z}^{k,l+1}&=\mathbf{e}_{cas}^{k,l}+\mathbf{z}^{k,l}+\sum_j\mathbf{w}^{k,l}(j) \cdot \mathbf{f}^{k,l}(j)
\end{aligned}
\end{array}\right.
\end{equation}
where $(\circ)$ is the hadamard product operation, $\mathbf{W}_{k,l} \in \mathbb{R}^{4 \times 4d}$ and $\mathbf{b}_{k,l} \in \mathbb{R}^{4 \times 1}$ are feature transformation matrix and bias matrix. $d$ is the dimension of embedding.

For the output of each behavior, we have:
\begin{equation}
% \left\{\begin{array}{c}
% \begin{aligned}
% \mathbf{p}^{k,*} &= \sum_{l=0}^{L_k}\mathbf{p}^{k,l}\\
% \mathbf{z}^{k,*} &= \sum_{l=0}^{L_k}\mathbf{z}^{k,l}
% \end{aligned}
% \end{array}\right.
\mathbf{p}^{k,*} = \sum_{l=0}^{L_k}\mathbf{p}^{k,l}, \mathbf{z}^{k,*} = \sum_{l=0}^{L_k}\mathbf{z}^{k,l}
\end{equation}
where $L_k$ is the number of GNN layers of the $k$-th behavior.
\subsection{Projection Disentangling Multi-Experts Network}
As we have obtained the representations of each behavior $k$ in the previous section, we need to design a proper structure to further leverage the multiplex signals with these representations. It has been verified in many methods \cite{mmoe,PLE,sharebottom} that a multi-task learning module can perfectly handle this. The MTL structure first couples the representations of all behaviors, then generate kinds of experts by the coupled input, further applies a gating mechanism to aggregate the expert information as the output, and finally utilizes the prediction losses of different behaviors to jointly optimize the model. 

However, the existing MTL structures utilize a couple representation as the input while introducing noise from other behaviors while using gating mechanisms to aggregate information from different experts. This leads to the gradient conflict during the learning process. Thus, we proposed a well-designed MTL module to handle the above problems. The following are details.

\subsubsection{Generating of Experts.}
As coupled input contains mixed information of different behaviors, making the gradient coupled and conflict, we do not combine the multi-behavioral representations together. We directly leverage each representation to generate the behavior-specific experts:
\begin{equation}
    \mathbf{q}^{k} = \mathbf{z}_u^{k,*} \circ \mathbf{z}_v^{k,*}
\end{equation}
where $(\circ)$ is the hadamard product operation.
\subsubsection{Aggregating of Experts.}
In order to alleviate the negative information transfer from other behavior-specific experts, we improve the gating mechanism with a representation projection mechanism. Take the behavior $k$ as an example, we have:
\begin{equation}
\setlength{\abovedisplayskip}{0.5pt}
\setlength{\belowdisplayskip}{0.5pt}
\left\{\begin{array}{c}
\begin{aligned}
\mathbf{q}_{sha}^{k^{\prime},k}&=\frac{\mathbf{q}^{k^{\prime}}\cdot\mathbf{q}^{k}}{|\mathbf{q}^{k}|} \frac{\mathbf{q}^{k}}{|\mathbf{q}^{k}|}\\
\mathbf{q}_{uni}^{k^{\prime},k}&=\mathbf{q}^{k^{\prime}}-\mathbf{q}_{sha}^{k^{\prime},k}
\end{aligned}
\end{array}\right.
\end{equation}
where ($\cdot$) is the vector inner product operation. $\mathbf{q}^{k^{\prime}}$ and $\mathbf{q}^{k}$ are the representations of the $k^{\prime}$- and $k$-th behavior, respectively. $\mathbf{q}_{sha}^{k^{\prime},k}$ contains a mixture of the $k^{\prime}$- and $k$-th behavioral correlation information. $\mathbf{q}_{uni}^{k^{\prime},k}$, which represents the unique part of the $k^{\prime}$-th behavior, and $\mathbf{q}^{k}$, which denotes the $k$-th behavior, are distinctive and should be as orthogonal as possible.

As we can see, the projection mechanism disentangle $\mathbf{q}^{k^{\prime}}$ by the guidance of $\mathbf{q}^{k}$, thus the shared and unique parts of other behaviors can be further utilized to alleviate the negative transfer caused by gating aggregation. To be specific, we take the shared representations $\mathbf{q}_{sha}^{k^{\prime},k} (k^{\prime} \in \{1,2,...,K\} \cap k^{\prime} \neq k)$ of other behaviors and $\mathbf{q}^{k}$ as targets of aggregation by the $k$-th gate, and have:
\begin{equation}
\left\{\begin{array}{c}
\begin{aligned}
\mathbf{g}^{k} &= Softmax(\mathbf{W}_{g}\mathbf{z}^{k,*} + \mathbf{b}_{g})\\
\hat{\mathbf{q}}^{k}&=\mathop{Concat}\limits_{k^{\prime}=1}^{K}(\mathbf{q}_{sha}^{k^{\prime},k})\\
\hat{o}_{uv}^{k,cas} &= h^{k}(\sum_{j=1}^{K} {\mathbf{g}^{k}(j) \cdot \hat{\mathbf{q}}^{k}(j)})
\end{aligned}
\end{array}\right.
\end{equation}
where $\mathbf{W}_{g} \in \mathbb{R}^{K \times d}$ and $\mathbf{b}_{g} \in \mathbb{R}^{K \times 1}$ are feature transformation matrix and bias matrix, and $\mathbf{g}^{k} \in \mathbb{R}^{K \times 1}$ is the attention vector which are used as selector to calculate the weighted sum of all experts. $h^{k}(\cdot)$ is the tower function. $\hat{o}_{uv}^{k,cas}$ is the prediction score of whether user $u$ will have interaction with item $v$ under behavior $k$ at the cascade stream.

Besides, we design a prediction task for the unique representation $\mathbf{q}_{uni}^{k^{\prime},k}$. This task takes full advantage of the mutually exclusive relationship between $\mathbf{q}_{uni}^{k^{\prime},k}$ and $\mathbf{q}^{k}$, facilitating the learning of the $k$-th behavior. Details are shown in Section \ref{uni_loss}.

\subsection{Joint Optimization}
\subsubsection{Parallel Loss.}
As we have obtained multi-behavioral representations from the parallel stream, we design a parallel loss to help the learning of each representations. In details, we have:
% \begin{equation}
% \left\{\begin{array}{c}
% \begin{aligned}
% \hat{o}_{uv}^{k,par} &= \mathbf{p}_u^{k,*} \cdot \mathbf{p}_v^{k,*} \\
% \mathcal{L}^{par} &= -\sum_{k=1}^{K}\sum_{(u,s,t)\in \mathcal{O}_k} \lambda_k * \textup{ln} \sigma(\hat{o}_{us}^{k,par} - \hat{o}_{ut}^{k,par})
% \end{aligned}
% \end{array}\right.
% \end{equation}
\begin{equation}
\mathcal{L}^{par} = -\sum_{k=1}^{K}\sum_{(u,s,t)\in \mathcal{O}_k} \lambda_k * \textup{ln} \sigma(\hat{o}_{us}^{k,par} - \hat{o}_{ut}^{k,par})
\end{equation}
where $\hat{o}_{uv}^{k,par}=\mathbf{p}_u^{k,*} \cdot \mathbf{p}_v^{k,*}$ is the prediction score of whether user $u$ will have interaction with item $v$ under behavior $k$ at the parallel stream. And we apply a \textit{Bayesian Personalized Ranking} (BPR) \cite{bpr} loss to optimize the model. $\mathcal{O}_k = \left\{(u,s,t)|(u,s)\in \mathcal{O}_k^{+}, (u,t) \in \mathcal{O}_k^{-} \right\}$ denotes the training dataset. $\mathcal{O}_k^+$ indicates observed positive user-item interactions under behavior $k$ and $\mathcal{O}_k^-$ indicates unobserved user-item interactions under behavior $k$. $\lambda_k$ is the coefficient of behavior $k$. $\sigma$ is the Sigmoid function.
\subsubsection{Cascade Loss.}
Similar to the above, we devise a cascade loss for the cascade stream, and as we have obtained the final prediction $\hat{o}_{uv}^{k,cas}$, we have:
\begin{equation}
\mathcal{L}^{cas} = -\sum_{k=1}^{K}\sum_{(u,s,t)\in \mathcal{O}_k} \lambda_k * \textup{ln} \sigma(\hat{o}_{us}^{k,cas} - \hat{o}_{ut}^{k,cas})
\end{equation}
where the definition of parameters is the same to what in the parallel loss.
\subsubsection{Unique Loss.}
\label{uni_loss}
To make full use of the unique representation $\mathbf{q}_{uni}^{k^{\prime},k}$, we design an auxiliary prediction task. Specifically, we leverage $\mathbf{q}_{uni}^{k^{\prime},k}$ to predict the interactive information of ``$k^{\prime}$ without $k$''. In details, we have:
% \begin{equation}
% \left\{\begin{array}{c}
% \begin{aligned}
% \hat{o}_{uv}^{k^{\prime},k,uni} &= \mathbf{q}_{u,uni}^{k^{\prime},k} \cdot \mathbf{q}_{v,uni}^{k^{\prime},k} \\
% \mathcal{L}^{uni} &= -\sum_{k=1}^{K}\sum\limits_{k^{\prime}=1 \atop k^{\prime} \neq k}^{K}\sum_{(u,s,t)\in \mathcal{T}_{k^{\prime},k}} \lambda_k * \textup{ln} \sigma(\hat{o}_{us}^{k^{\prime},k,uni} - \hat{o}_{ut}^{k^{\prime},k,uni})
% \end{aligned}
% \end{array}\right.
% \end{equation}
\begin{equation}
\mathcal{L}^{uni} = -\sum_{k=1}^{K}\sum\limits_{k^{\prime}=1 \atop k^{\prime} \neq k}^{K}\sum_{(u,s,t)\in \mathcal{T}_{k^{\prime},k}} \lambda_k * \textup{ln} \sigma(\hat{o}_{us}^{k^{\prime},k,uni} - \hat{o}_{ut}^{k^{\prime},k,uni})
\end{equation}
where $\hat{o}_{uv}^{k^{\prime},k,uni}=\mathbf{q}_{u,uni}^{k^{\prime},k} \cdot \mathbf{q}_{v,uni}^{k^{\prime},k}$ is the prediction score. $\mathcal{T}_{k^{\prime},k} = \left\{(u,s,t)|(u,s)\in \mathcal{O}_{k^{\prime}}^{+} \cap \mathcal{O}_k^{-}, (u,t) \in \mathcal{O}_k^{-} \cup (\mathcal{O}_{k^{\prime}}^{+} \cap \mathcal{O}_k^{+}) \right\}$ denotes the training dataset. $\mathcal{O}_k^+$ indicates observed positive user-item interactions under behavior $k$ and $\mathcal{O}_k^-$ indicates unobserved user-item interactions under behavior $k$. In short, we remove from the behavioral adjacency matrix $\mathbf{M}_{k^{\prime}}$ the positive items that $\mathbf{M}_{k^{\prime}}$ shares with $\mathbf{M}_{k}$.
% we change the positive items that have interacted with the same user under $k$ behavior in the adjacency matrix of behavior $k^{\prime}$ to negative items. 
Thus, we fully utilize the "Only $k^{\prime}$" interactive information with the help of the unique representations of behavior $k^{\prime}$.

In all, the final loss can be formulated as:
\begin{equation}
    \mathcal{L}(\Theta) = \mathcal{L}^{par} + \mathcal{L}^{cas} + \mathcal{L}^{uni} + \mu ||\Theta||^2_2
\end{equation}
where $\Theta$ represents set of all model parameters. $\mu$ is the $L_2$ regularization coefficient for $\Theta$.
\subsection{Complexity Analysis}
\label{complexity}
\subsubsection{Time Complexity.}
The time complexity of PKEF primarily lies in the GNN parts, which consist of cascade and parallel streams. Both the cascade and parallel parts have a computational complexity of $\sum_{k=1}^{K}{O\left(L^{k}\cdot\left|\mathcal{E}_{k} \right| \cdot d\right)}$. Here, $\left|\mathcal{E}_{k} \right|$ represents the number of edges across all graphs in the set $\mathcal{E}_{k}$, $K$ denotes the behavior number, $L_k$ refers to the number of GNN layers of the $k$-th behavior, and $d$ represents the embedding size.
Overall, the time complexity of PKEF is comparable to that of existing GNN-based methods.
\subsubsection{Space Complexity.} 
The learnable parameters in our proposed PKEF primarily come from the user and item embeddings, denoted as $\mathbf{x}_u$ and $\mathbf{y}_i$ respectively. This is similar to existing methods.
Furthermore, the dense graphs $\mathcal{G}_{k}$ in the set $\mathcal{G}$ are transformed into sparse behavior-specified matrices $\mathbf{M}_1, \mathbf{M}_2, \cdots, \mathbf{M}_K$ for computational purposes. This transformation allows us to perform computations without requiring additional memory space to store the dense graphs. Hence, the memory usage during the intermediate process remains within an acceptable range.

\section{Experiments}
\label{experiments}

\begin{table}[t]
\setlength{\abovecaptionskip}{0cm}
\setlength{\belowcaptionskip}{0mm}
\caption{Statistics of evaluation datasets.}
\centering
\resizebox{\linewidth}{!}{
\begin{tabular}{c|ccccc}
\toprule
Dataset & \#User & \#Item & \#Interaction & \#Target Interaction &  \#Interactive Behavior Type \\ \hline
Beibei & 21,716 & 7,977 & $3.3 \times 10^6$ & 282,860 & \{View,Cart,Buy\} \\ \hline
Taobao & 15,449 & 11,953 & $1.2 \times 10^6$ & 92,180 & \{View,Cart,Buy\} \\ \hline
Tmall & 41,738 & 11,953 & $2.3 \times 10^6$ & 255,586 & \{View,Collect,Cart,Buy\} \\  \bottomrule

\end{tabular}
}
\vspace{-4mm}
\label{tab:dataset}
\end{table}

\subsection{Experimental Setting}
\subsubsection{Dataset Description}
We follow MB-CGCN \cite{mbcgcn} and CRGCN \cite{crgcn}, and adopt the same three datasets for evaluation, i.e., Beibei, Taobao and Tmall. For these datasets, we adhere to previous studies' methodology of removing duplicates by retaining the earliest entry \cite{mbgcn,gao2019learning}. Table \ref{tab:dataset} provides a summary of the statistical information for the three datasets used in our experiments.

\subsubsection{Evaluation Protocols}
In all our experiments, we assess the performance of our proposed PKEF model and baseline models based on the top-$k$ recommended items, using two evaluation metrics: Hit Ratio (\textit{HR@k}) and Normalized Discounted Cumulative Gain (\textit{NDCG@k}). Specifically, we set $k = 10$ for our evaluations.

\subsubsection{Baseline Models}
To demonstrate the effectiveness of PKEF, we compare it with several state-of-the-art methods, which can be divided into two categories: \textbf{(1) Single-behavior methods:} MF-BPR \cite{bpr}, NeuMF \cite{ncf} and LightGCN\footnote{https://github.com/kuandeng/LightGCN} \cite{lightgcn}, \textbf{(2) Multi-behavior methods without MTL:} RGCN \cite{RGNN}, GNMR \cite{gnmr}, NMTR \cite{nmtr}, \\ MBGCN\footnote{https://github.com/tsinghua-fib-lab/MBGCN} \cite{mbgcn}, CRGCN\footnote{https://github.com/MingshiYan/CRGCN} \cite{crgcn} and MB-CGCN\footnote{https://github.com/SS-00-SS/MBCGCN} \cite{mbcgcn},

\subsubsection{Parameter Settings}
Our proposed PKEF is implemented in TensorFlow \cite{TensorFlow}. For a fair comparison, following MB-CGCN \cite{mbcgcn}, we set the embedding size to 64. We initialize the parameters using Xavier \cite{xavier}. The parameters are optimized by Adam \cite{adam}, while the learning rate is set to $10^{-3}$. We search the number of GNN layers for every behavior in \{1,2,3,4\} for user-item bipartite graph. In addition, we adjust the loss coefficients for each behavior in \{0,1/6,2/6,3/6,4/6,5/6,1\} and fix the sum of the coefficients for all actions as 1. Other parameters are the same as MB-CGCN. All experiments are run for 5 times and average results are reported.
% Since the existing methods compare the performance of models with two mainstream settings, 
In addition, we conduct hyper-parameter analysis experiments (shown in Section \ref{hyper}).

\begin{table}[t]
\setlength{\abovecaptionskip}{0cm}
\setlength{\belowcaptionskip}{0mm}
\caption{The overall performance comparison. Boldface denotes the highest score and underline indicates the results of the best baselines. $\star$ represents significance level $p$-value $<0.05$ of comparing PKEF with the best baseline.}
    \centering
    \begin{threeparttable}
	\resizebox{\linewidth}{!}{
    \begin{tabular}{c|cccccc}
    \toprule
    \multirow{2}{*}{Model}&
    \multicolumn{2}{c}{Beibei}&\multicolumn{2}{c}{Taobao}&\multicolumn{2}{c}{Tmall}\cr
    \cmidrule(lr){2-3} \cmidrule(lr){4-5} \cmidrule(lr){6-7} 
    &HR&NDCG&HR&NDCG&HR&NDCG\cr
    \midrule
    MF-BPR&0.0191&0.0049&0.0076&0.0036&0.0230&0.0207\cr
    NeuMF&0.0232&0.0135&0.0236&0.0128&0.0124&0.0062\cr
    LightGCN&0.0391&0.0209&0.0411&0.0240&0.0393&0.0209\cr
    \hline
    RGCN&0.0363&0.0188&0.0215&0.0104&0.0316&0.0157\cr
    GNMR&0.0413&0.0221&0.0368&0.0216&0.0393&0.0193\cr
    NMTR&0.0429&0.0198&0.0282&0.0137&0.0536&0.0286\cr
    MBGCN&0.0470&0.0259&0.0509&0.0294&0.0549&0.0285\cr
    CRGCN&0.0459&0.0324&0.0855&0.0439&0.0840&0.0442\cr
    MB-CGCN&\underline{0.0579}&\underline{0.0381}&\underline{0.1233}&\underline{0.0677}&\underline{0.0984}&\underline{0.0558}\cr
    \hline
    \textbf{PKEF}&\textbf{0.1130}$^\star$&\textbf{0.0582}$^\star$&\textbf{0.1385}$^\star$&\textbf{0.0785}$^\star$&\textbf{0.1277}$^\star$&\textbf{0.0721}$^\star$\cr
    \hline
    Rel Impr.&95.16\%&38.58\%&12.33\%&15.95\%&29.78\%&29.21\%\cr
    \bottomrule
    \end{tabular}}
    \end{threeparttable}
    \vspace{-4mm}
    \label{comparisons_model}
\end{table}

\subsection{Performance Comparison}
\label{performance}
Table \ref{comparisons_model} shows the performance of methods on three datasets with respect to HR@10 and NDCG@10. 
We have the following findings:
\begin{itemize}

\item Our PKEF model achieves the best performance across all three datasets. Specifically, in terms of HR and NDCG metrics, PKEF outperforms the best baselines on Beibei, Taobao, and Tmall datasets by \textbf{95.16$\%$}, \textbf{12.33$\%$},\textbf{ 29.78$\%$} and \textbf{38.58$\%$}, \textbf{15.95$\%$}, \textbf{29.21$\%$}, respectively. Our PKEF model demonstrates significant enhancements in recommendation accuracy, particularly when compared to the best baseline, MB-CGCN. This substantial progress highlights the effectiveness of our model.

\item Multi-behavior models perform better than single-behavior models. For example, MBGCN performs better than LightGCN. This indicates the superiority of utilizing multiple types of interactions.

\item LightGCN consistently outperforms MF-BPR and NeuMF, while MBGCN outperforms NMTR. This demonstrates the advantages of the GCN model, which leverages high-order neighbor information on the user-item bipartite graph to learn embeddings for users and items.

\item Finally, GNMR and MBGCN outperform RGCN by considering the contribution of each behavior in the multi-behavioral fusion step. Compared to NMTR and MBGCN, which only propose parallel learning paradigms during behavior fusion, CRGCN and MB-CGCN explicitly incorporate the cascade relationships of multiple behaviors during the fusion step, achieving performance that is second only to our model. This indicates the necessity of considering hierarchical correlation between behaviors.
\end{itemize}

\subsection{Ablation Study}
\label{ablation_study}
\subsubsection{Impact of the Key Components}

\begin{table}[t]
    \setlength{\abovecaptionskip}{0cm}
    \setlength{\belowcaptionskip}{0mm}
    \caption{Performances of different PKEF variants.}
    \centering
    \begin{threeparttable}
    \resizebox{\linewidth}{!}{
    \begin{tabular}{c|cccccc}
    \toprule
    \multirow{2}{*}{Model}&
    \multicolumn{2}{c}{Beibei}&\multicolumn{2}{c}{Taobao}&\multicolumn{2}{c}{Tmall}\cr
    \cmidrule(lr){2-3} \cmidrule(lr){4-5} \cmidrule(lr){6-7}
    &HR&NDCG&HR&NDCG&HR&NDCG\cr
    \midrule
    Base Model &0.0734&0.0360&0.0849&0.0452&0.0829&0.0466\cr
    PKEF w/o PKF &0.1096&0.0560&0.1121&0.0610&0.1205&0.0687\cr
    % PKEF w/o PME &0.0940&0.0473&0.1064&0.0596&0.1115&0.0639\cr
    PKEF w/o PME &0.0915&0.0460&0.0955&0.0532&0.0931&0.0514\cr
    PKEF &\textbf{0.1130}  &\textbf{0.0582}  &\textbf{0.1385}  &\textbf{0.0785}  &\textbf{0.1277} &\textbf{0.0721}\cr
    \bottomrule
    \end{tabular}}
    \end{threeparttable}
    \vspace{-4mm}
    \label{tab:ablation_key}
\end{table}
To evaluate the effectiveness of sub-modules in our PKEF framework, we consider three model variants: (1) \textbf{Base Model}: We remove both the PKF and PME parts, so that the model only has the cascade stream and utilizing a bilinear paradigm (separated input with a light-weight matrix transformation); (2) \textbf{PKEF w/o PKF}: The PKF part is removed; (3) \textbf{PKEF w/o PME}: The PME part is replaced with bilinear module.
The performance of PKEF and its variants are summarized in Table \ref{tab:ablation_key}, and we come to these conclusions:
\begin{itemize}
\item Comparing the performance of PKEF and its last two variants, we can find that each variant brings about performance degradation when any key component is removed or replaced with other modules.
This demonstrates the rationality and effectiveness of the two key designations.
\item It is worthwhile noticing that Base Model achieves the worst performance on all three datasets compared to other variants with multi-behavior learning.
In particular, this variant has a performance decline up to \textbf{35.04\%}, \textbf{38.70\%}, and \textbf{35.08\%} in terms of HR (\textbf{38.14\%}, \textbf{42.42\%}, and \textbf{35.37\%} in terms of NDCG) on Beibei, Taobao, and Tmall datasets.
This further demonstrates the effectiveness of the combination of PKF and PME for solving the multi-behavior recommendation problem.
\end{itemize}

\subsubsection{Impact of the Knowledge Fusion Schemes}
\label{knowledge_fusion}
\begin{table}[t]
    \setlength{\abovecaptionskip}{0cm}
    \setlength{\belowcaptionskip}{0mm}
    \caption{Performances of different knowledge fusion schemes.}
    \centering
    \begin{threeparttable}
    \resizebox{\linewidth}{!}{
    \begin{tabular}{c|cccccc}
    \toprule
    \multirow{2}{*}{Model}&
    \multicolumn{2}{c}{Beibei}&\multicolumn{2}{c}{Taobao}&\multicolumn{2}{c}{Tmall}\cr
    \cmidrule(lr){2-3} \cmidrule(lr){4-5} \cmidrule(lr){6-7}
    &HR&NDCG&HR&NDCG&HR&NDCG\cr
    \midrule
    Summation &0.0312&0.0173&0.1142&0.0635&0.0238&0.0127\cr
    Linear Trans.&0.0847&0.0419&0.1185&0.0666&0.1084&0.0617\cr
    Vanilla Fusion &0.1105&0.0568&0.1254&0.0707&0.1180&0.0674\cr
    Projection Fusion&\textbf{0.1130}  &\textbf{0.0582}  &\textbf{0.1385}  &\textbf{0.0785}  &\textbf{0.1277} &\textbf{0.0721}\cr
    \bottomrule
    \end{tabular}}
    \end{threeparttable}
    \vspace{-4mm}
    \label{tab:ablation_fusion}
\end{table}

To further explore the forms of knowledge fusion between the parallel and cascade streams, we devise two alternative schemes for general usage (Illustrated in Section \ref{para_inter_enhance}). Besides, we make a comparison between the proposed two schemes with simple \textbf{Summation} (simply add the representation of different streams up) and \textbf{Linear Trans.} (apply a linear transformation to transfer the parallel knowledge).
And as shown in Table \ref{knowledge_fusion}, we can observe that summation perform the worst among the four schemes. A probable reason is that the distribution of the two representations of the streams is completely different. So, simple summation may cause the harmful impact to the distribution of representations. Besides, Linear Trans. leverage a implicit way to transfer the knowledge, which may lead to a negative information transfer when transfering the parallel knowledge. Vanilla Fusion weights the fusion representations at different scales, alleviating the impact of representation distribution. While the Projection Fusion utilizes a projection mechanism to explicitly extract the useful information from the parallel knowledge, and thus obtain the best performance on these three datasets.

\subsubsection{Impact of the MTL module}
To further demonstrate the superiority of our proposed PME in Multi-Task Learning (MTL), we compare it with four state-of-the-art MTL models: Shared Bottom \cite{sharebottom}, Bilinear \cite{ghcf}, MMOE \cite{mmoe}, and PLE \cite{PLE}. These models are applied on top of PKF for multi-behavior recommendation.
To ensure compatibility with the classical MTL models (i.e., Shared Bottom, MMOE, and PLE), which expect the same input representation, we weigh the $K$ separate representations generated by PKF to obtain a unified input. The resulting variants are named PKF+SB, PKF+Bilinear, PKF+MMOE, PKF+PLE, and PKF+PME.
Table \ref{tab:ablation_mtl} summarizes the results. PKF+SB performs the worst among all MTL models across all datasets. PKF+Bilinear, which replaces the neural network's prediction head with a light-weight matrix transformation, shows better performance, which is likely due to reduced risk of overfitting.
Both PKF+MMOE and PKF+PLE employ gate networks with adaptive attention weights for information fusion, outperforming the static and equally weighted PKF+SB. Notably, our PME consistently outperforms all other models on all datasets, reaffirming its effectiveness for MTL tasks.

\begin{table}[t]
    \setlength{\abovecaptionskip}{0cm}
    \setlength{\belowcaptionskip}{0mm}
    \caption{Performances of different MTL module.}
    \centering
    \begin{threeparttable}
    \resizebox{\linewidth}{!}{
    \begin{tabular}{c|cccccc}
    \toprule
    \multirow{2}{*}{Model}&
    \multicolumn{2}{c}{Beibei}&\multicolumn{2}{c}{Taobao}&\multicolumn{2}{c}{Tmall}\cr
    \cmidrule(lr){2-3} \cmidrule(lr){4-5} \cmidrule(lr){6-7}
    &HR&NDCG&HR&NDCG&HR&NDCG\cr
    \midrule
    PKF+SB &0.0599&0.0287&0.0715&0.0381&0.0840&0.0472\cr
    PKF+Bilinear &\underline{0.0915}&\underline{0.0460}&0.0955&0.0532&0.0931&0.0514\cr
    PKF+MMOE &0.0830&0.0420&\underline{0.1091}&\underline{0.0580}&0.0890&0.0487\cr
    PKF+PLE &0.0845&0.0431&0.1022&0.0545&\underline{0.0944}&\underline{0.0517}\cr
    PKF+PME &\textbf{0.1130}  &\textbf{0.0582}  &\textbf{0.1385}  &\textbf{0.0785}  &\textbf{0.1277} &\textbf{0.0721}\cr
    \bottomrule
    \end{tabular}}
    \end{threeparttable}
    \vspace{-4mm}
    \label{tab:ablation_mtl}
\end{table}

\begin{figure}[t]
	\setlength{\belowcaptionskip}{0cm}
	\setlength{\abovecaptionskip}{0cm}
	\subfigure{
        \begin{minipage}[t]{0.47\linewidth}
        \centering
		\label{fig:layer_beibei} 
		\includegraphics[width=\textwidth]{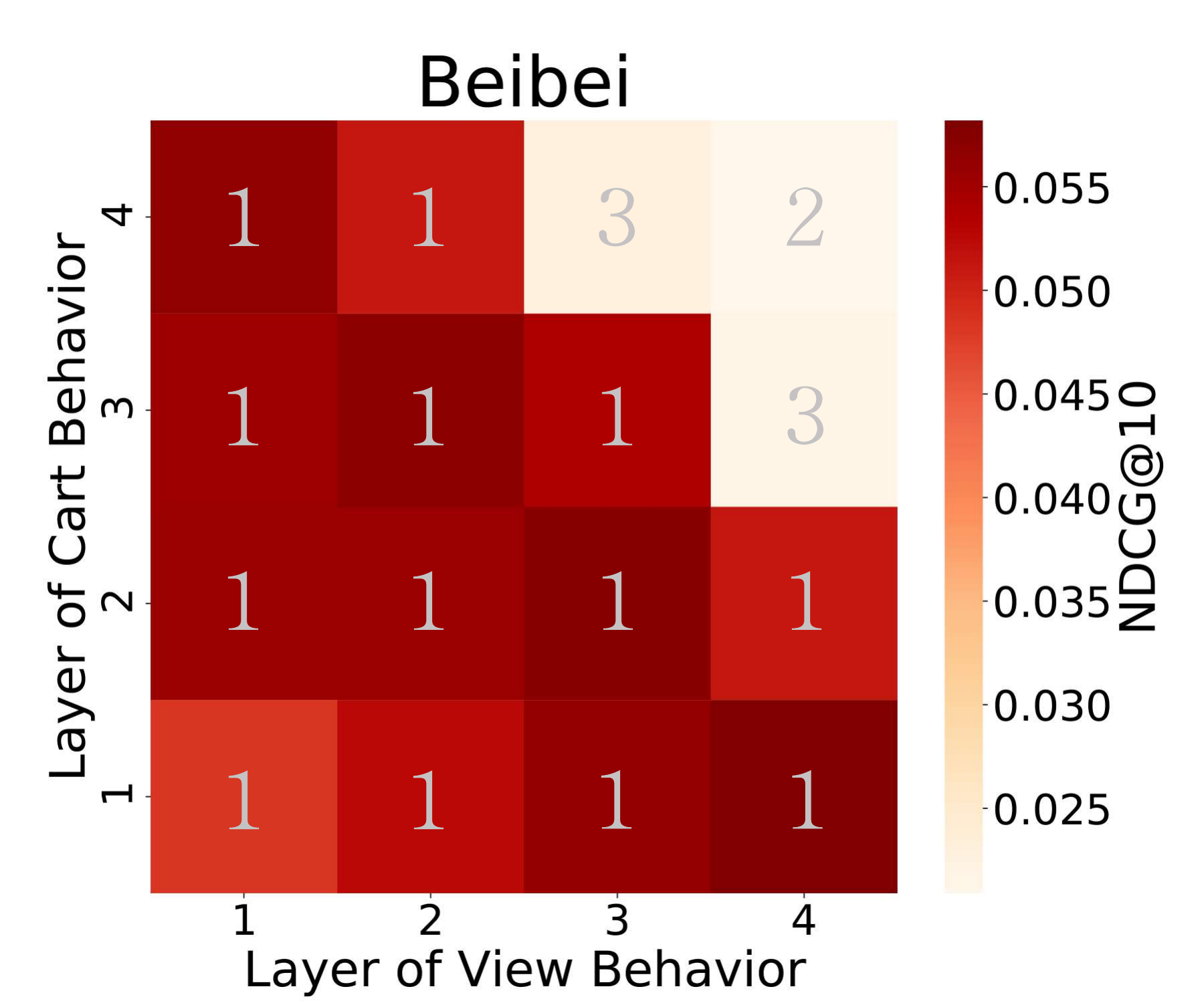}
        \end{minipage}}
	\subfigure{
        \begin{minipage}[t]{0.47\linewidth}
        \centering
		\label{fig:layer_taobao} 
		\includegraphics[width=\textwidth]{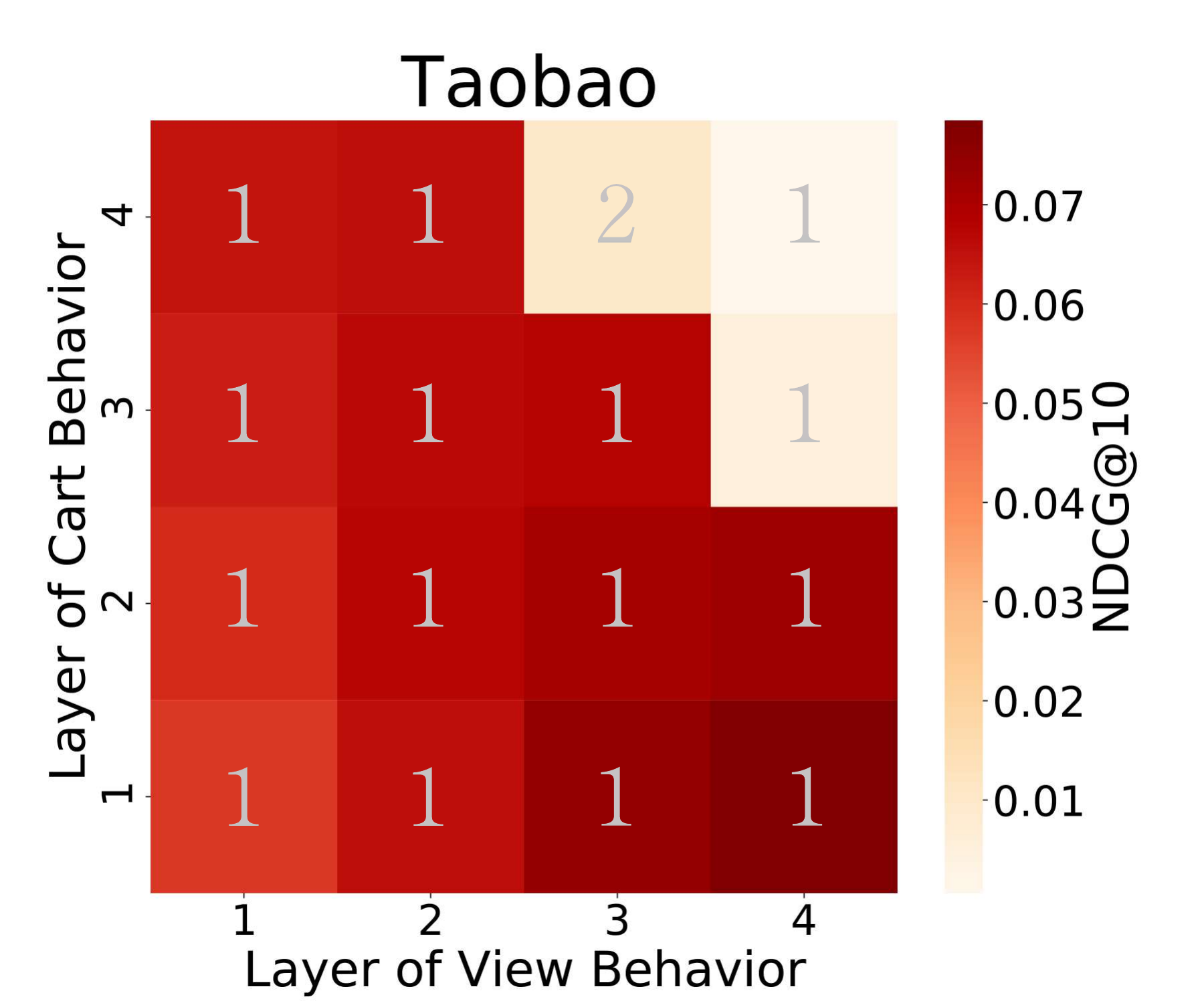}
        \end{minipage}}
	\caption{Impact of GNN Layers for different behaviors.}
        \vspace{-3mm}
	\label{fig:layer}
\end{figure}

\subsection{Parameter Analysis}
\label{hyper}
\subsubsection{Impact of the number of layers}
We investigate the impact of higher-order information on model performance by varying the number of GNN layers. Specifically, we search the layer numbers in the range of \{1, 2, 3, 4\} and use different numbers of layers for different behaviors. The experimental results are shown in Figure \ref{fig:layer}, where the numbers on each block indicate the layer for the \textit{buy} behavior that achieves the best performance while keeping the layer numbers fixed for the \textit{view} and \textit{cart} behaviors.
Due to lack of space, we only show the results on Beibei and Taobao, the results of another dataset are similar.

Based on the results, it is evident that for both datasets, PKEF demonstrates the highest performance when the GNN layers are configured as (4, 1, 1). Furthermore, the influence of stacking different numbers of layers on performance varied for different behaviors. There is a tendency to use deeper propagation layers for the graph of \textit{view} and shallower layers for downstream behaviors such as \textit{buy}. One possible reason is that the \textit{view} behavior contains richer interaction information and requires stacking more layers to capture higher-order information for learning better user preferences. Whereas in downstream behaviors with sparse interactions, excessive layers may introduce noise and lead to overfitting.

\subsubsection{Impact of the coefficients of different behaviors}
\begin{figure}[t]
	\setlength{\belowcaptionskip}{0cm}
	\setlength{\abovecaptionskip}{0cm}
	\subfigure{
        \begin{minipage}[t]{0.47\linewidth}
        \centering
		\label{fig:coefficient_beibei} 
		\includegraphics[width=\textwidth]{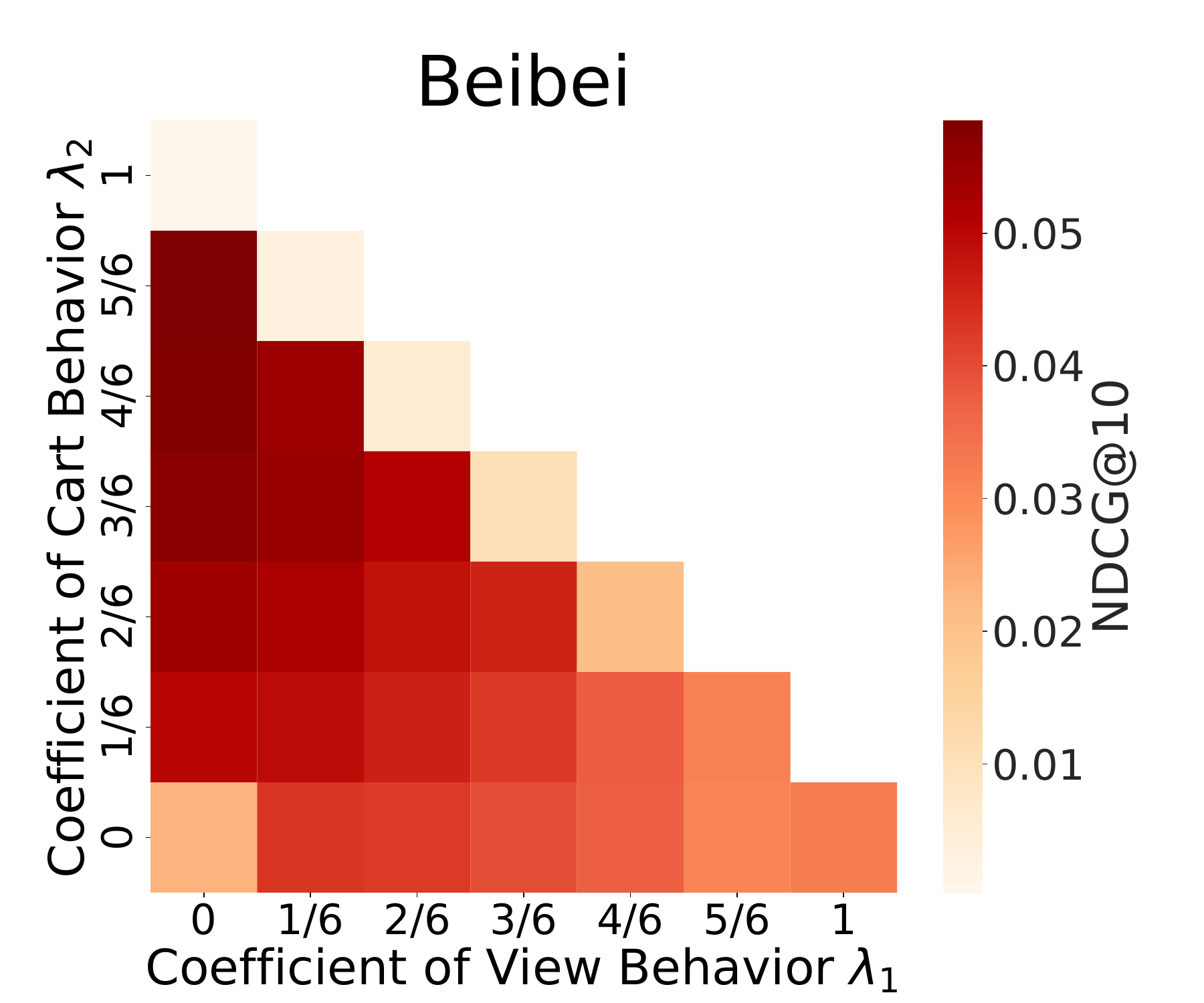}
        \end{minipage}}
	\subfigure{
        \begin{minipage}[t]{0.47\linewidth}
        \centering
		\label{fig:coefficient_taobao} 
		\includegraphics[width=\textwidth]{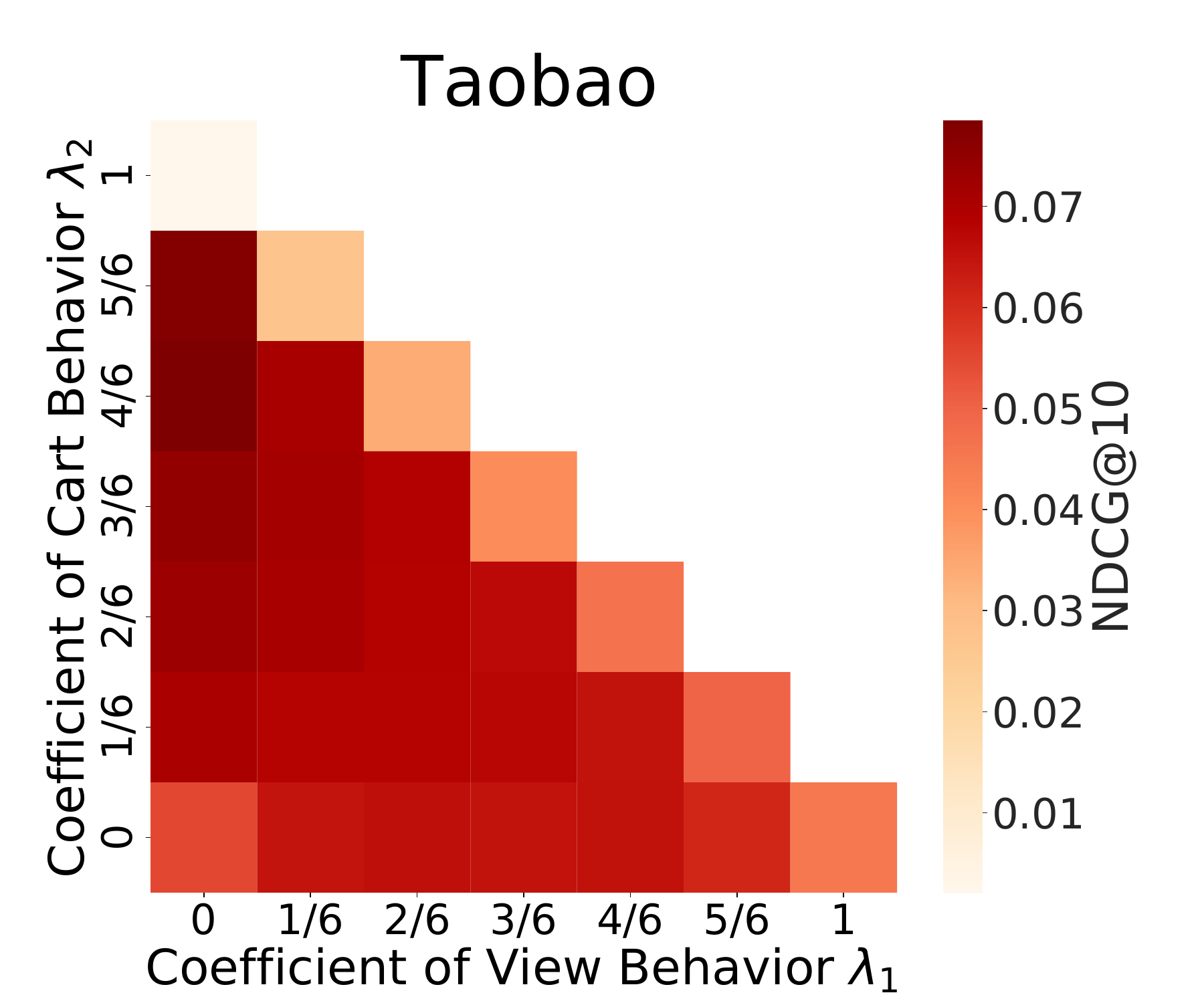}
        \end{minipage}}
	\caption{Impact of the Behavioral Coefficients.}
        \vspace{-3mm}
	\label{fig:coefficient}
\end{figure}

We investigate the impact of the behavioral coefficient parameter $\lambda_{k}$ on the performance of PKEF. There are three behavior types in Beibei and Taobao (\textit{view}, \textit{cart}, and \textit{buy}), which means there are three loss coefficients $\lambda_{1}$, $\lambda_{2}$, and $\lambda_{3}$, respectively. The value of $\lambda_{3}$ is determined when $\lambda_{1}$ and $\lambda_{2}$ are given. We use grid search in the range \{0, 1/6, 2/6, 3/6, 4/6, 5/6, 1\} and plot the results for NDCG@10 (shown in Figure \ref{fig:coefficient}). For both datasets, PKEF achieves the best performance with coefficient parameters set to (0, 4/6, 2/6), and the performance remains relatively consistent across different parameters. This indicates that the model can effectively adapt to different data distributions and has good generalization ability. The results of the Tmall dataset, which we omit due to space constraints, reach similar conclusions.

\subsection{Indepth Analysis}

\subsubsection{Case Study under Different Behavioral Correlations}
\begin{figure}[t]
	\setlength{\belowcaptionskip}{0cm}
	\setlength{\abovecaptionskip}{0cm}
	\subfigure{
        \begin{minipage}[t]{0.47\linewidth}
        \centering
		\label{fig:case_beibei} 
		\includegraphics[width=\textwidth]{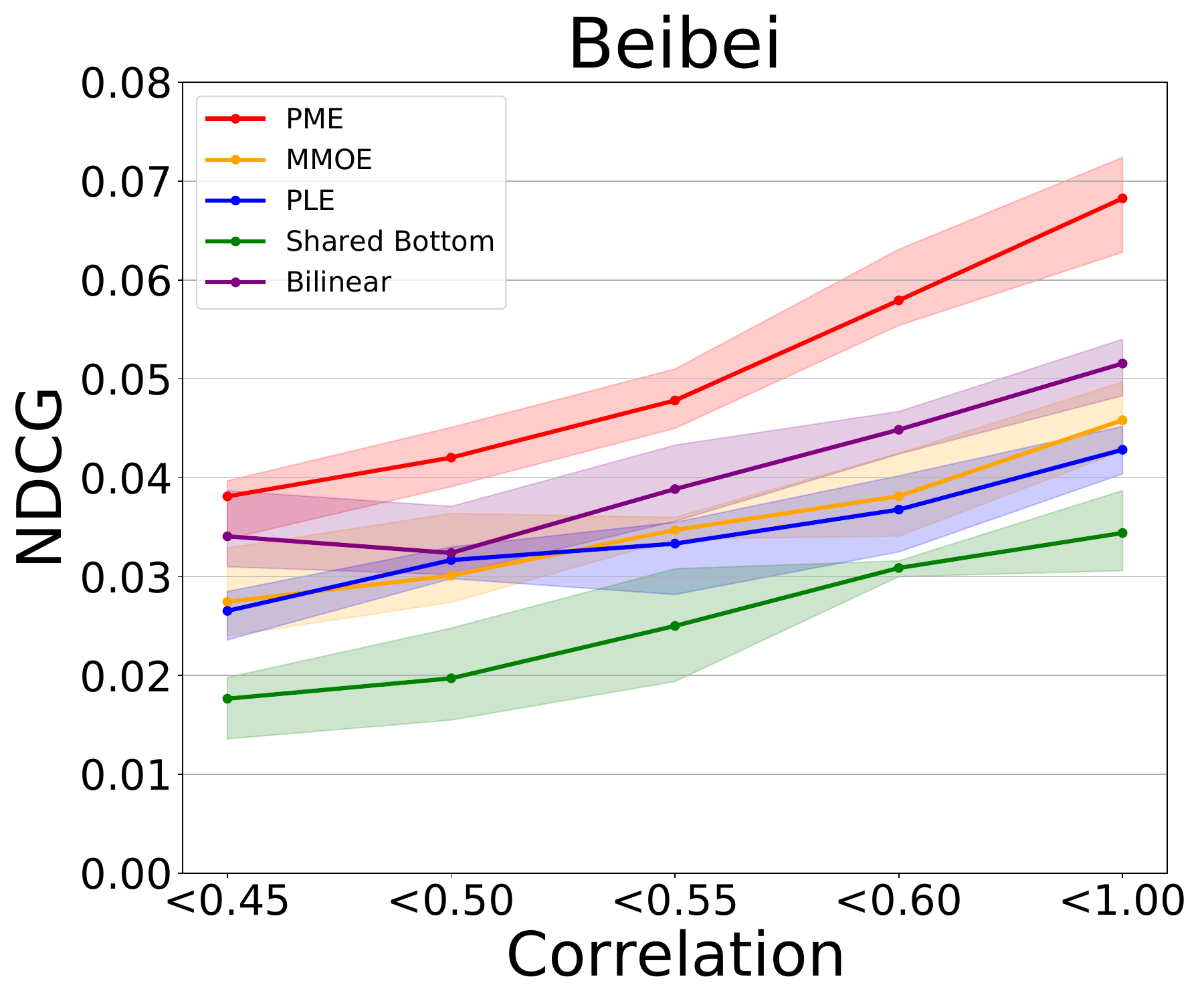}
        \end{minipage}}
	\subfigure{
        \begin{minipage}[t]{0.47\linewidth}
        \centering
		\label{fig:case_taobao} 
		\includegraphics[width=\textwidth]{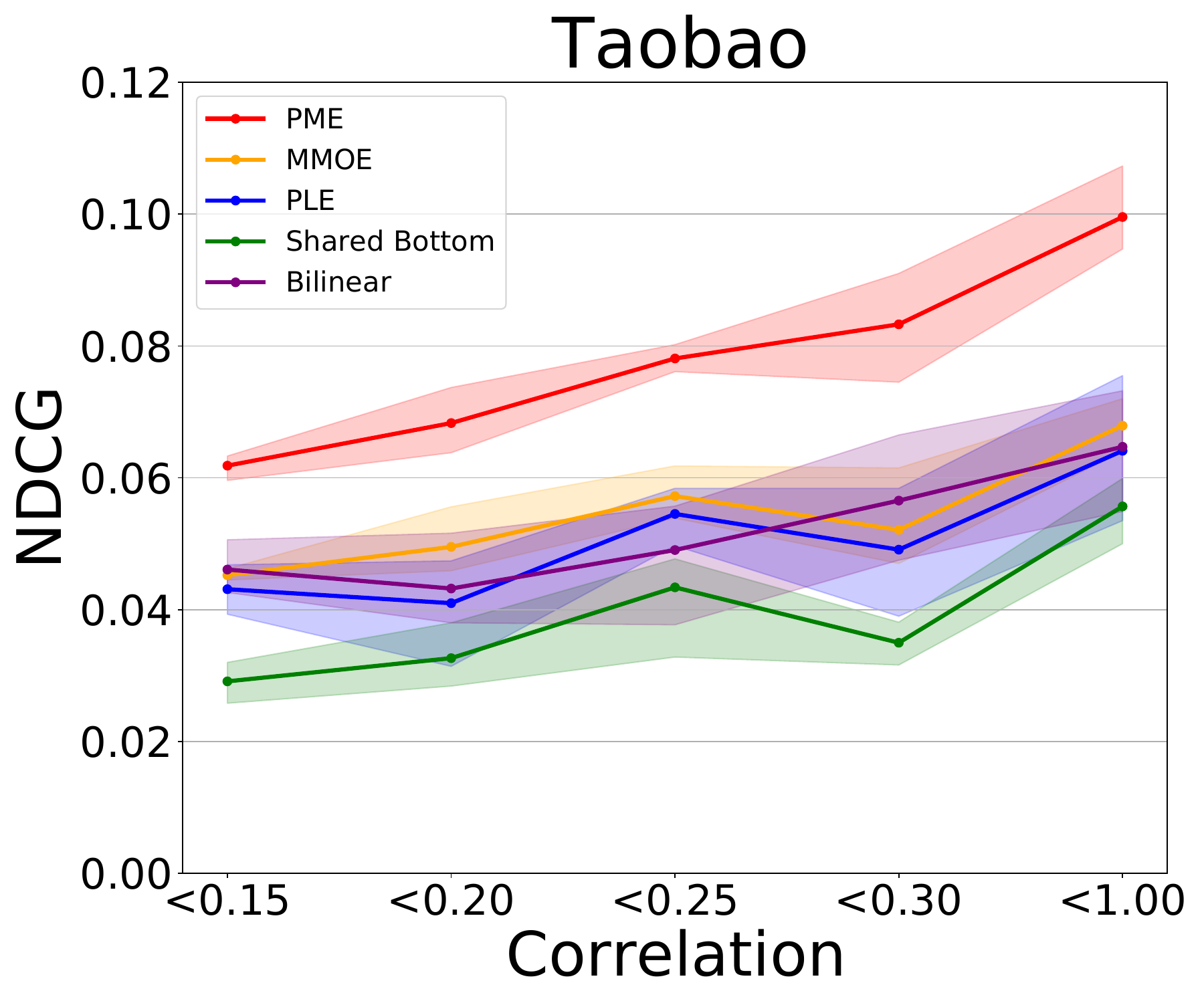}
        \end{minipage}}
	\caption{Average performances for user groups with different behavior correlations.}
        \vspace{-3mm}
	\label{fig:case_study}
\end{figure}

We experimentally verify whether our model can alleviate potential gradient conflicts. Specifically, we divide the test users into five user groups according to the average Pearson correlation among all behaviors and select subsets from each user group. To prevent the node degree from potentially influencing the results \cite{ngcf}, we keep similar average number of user interactions among different subsets while maximizing the number of users in each subset. For more rigorous results, we run experiments 5 times on each dataset and plot the mean and fluctuation range on the figure. The experimental results on Beibei and Taobao datasets are shown in Figure \ref{fig:case_study}. We find that PME consistently outperforms all other MTL methods across all user groups, further demonstrating the superiority of PME for MTL. Additionally, with the increase of Pearson correlation, the performance of PME grows more rapidly compared to other MTL methods, while other MTL methods even show fluctuations and decline. A possible reason is the negative transfer caused by potential gradient conflicts when knowledge is transferred across different tasks. We omit the results on the Tmall dataset due to space limitations, which have consistent conclusions.

\subsubsection{Visualization of Gating Aggregation}
\begin{figure}[t]
    \setlength{\belowcaptionskip}{0cm}
    \setlength{\abovecaptionskip}{0cm}
    \subfigure{
        \begin{minipage}[t]{0.3\linewidth}
        \centering
        \label{fig:pme} 
        \includegraphics[width=1in]{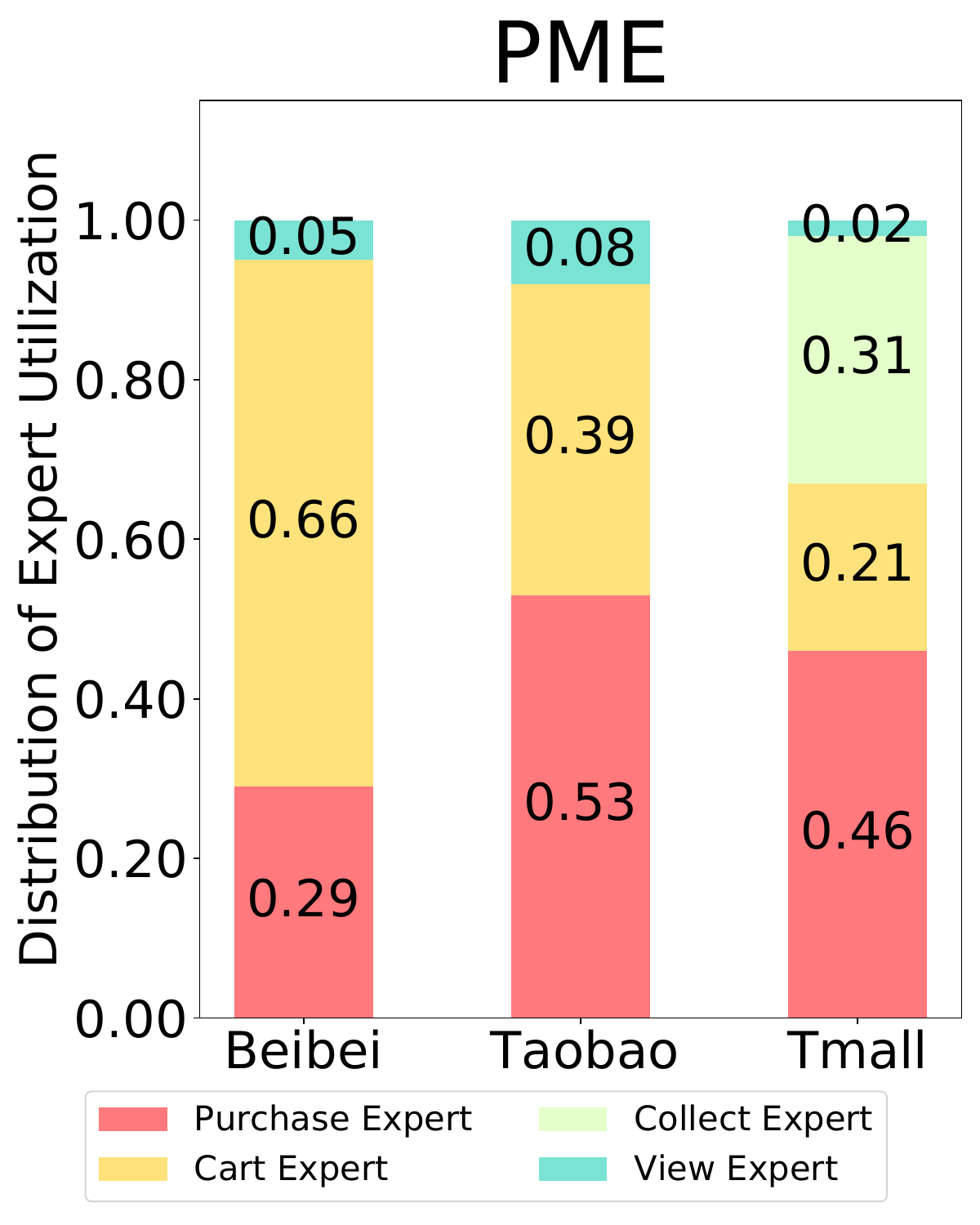}
        \end{minipage}}
    \subfigure{
        \begin{minipage}[t]{0.3\linewidth}
        \centering
        \label{fig:mmoe} 
        \includegraphics[width=1in]{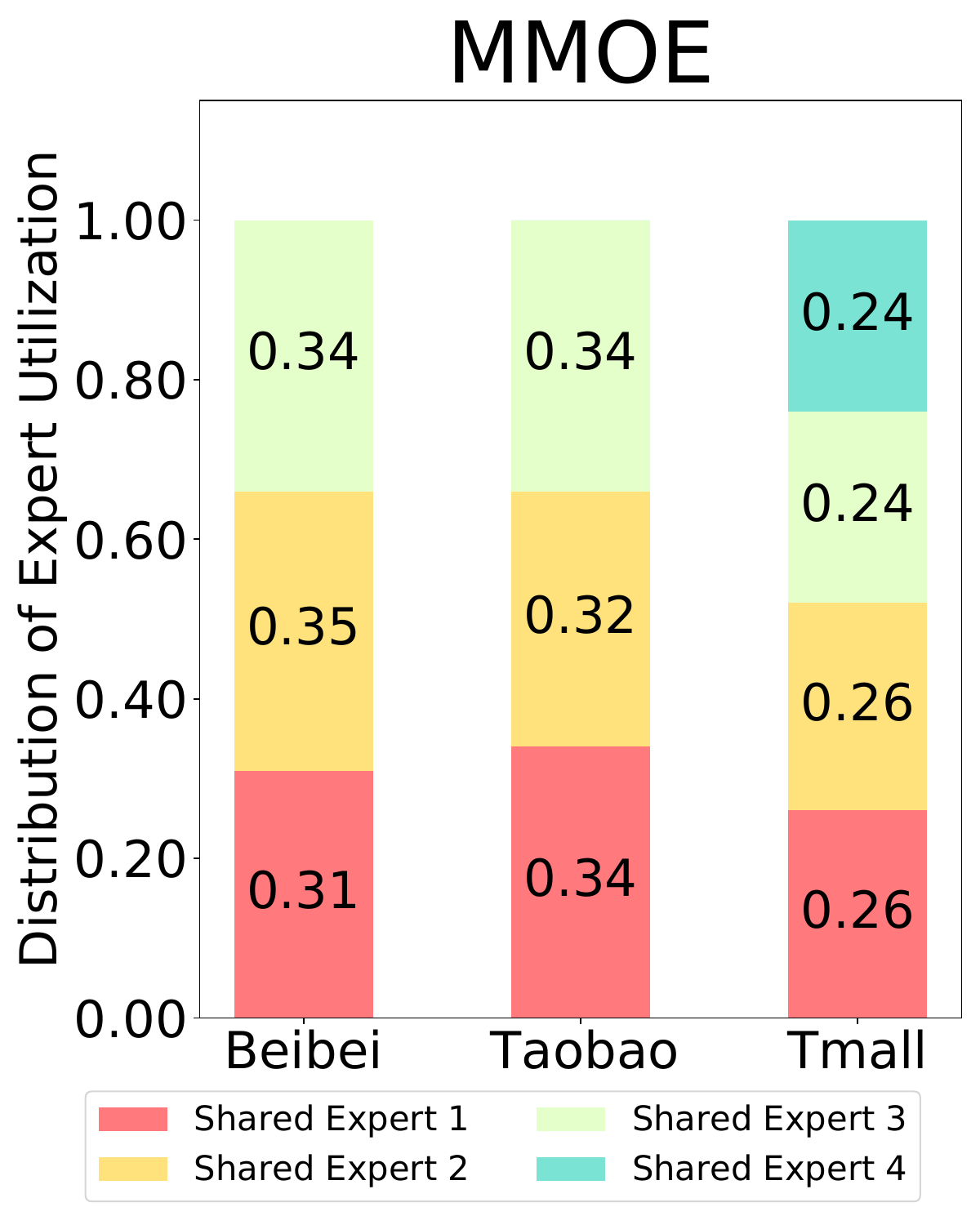}
        \end{minipage}}
    \subfigure{
        \begin{minipage}[t]{0.3\linewidth}
        \centering
        \label{fig:ple} 
        \includegraphics[width=1in]{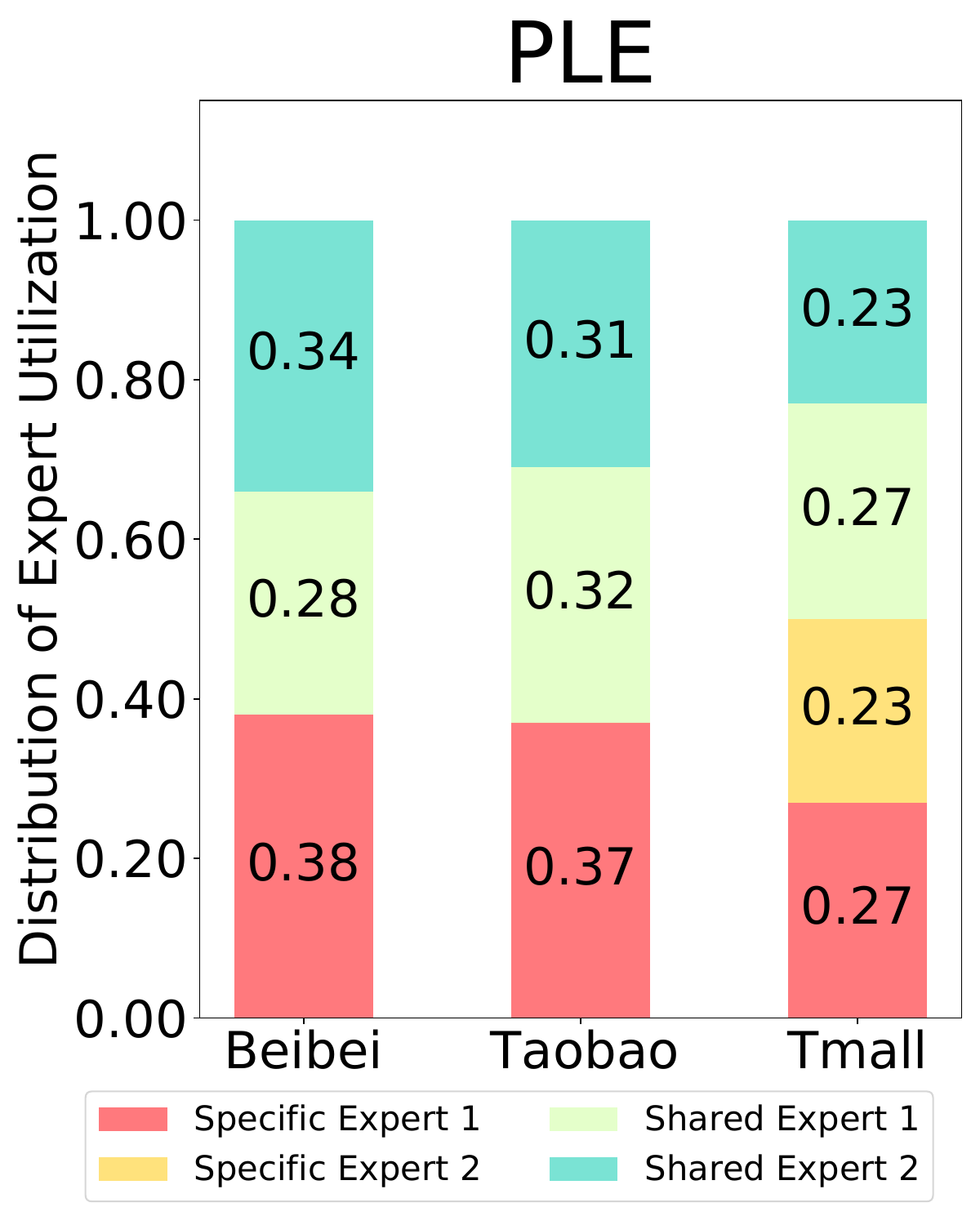}
        \end{minipage}}
    \Description{Expert utilization in gate-based models}
    \caption{Expert utilization in gate-based models}
    \vspace{-3mm}
    \label{fig:gate_distribution}
\end{figure}

We conduct experiments to compare the expert utilization between our PME model and other gate-based models (MMOE and PLE). By visualizing the average weight distribution of experts used for predicting the target behavior (shown in Figure \ref{fig:gate_distribution}), we observe that PME achieves better differentiation among experts compared to MMOE and PLE. 
We exclude gates used for other behaviors in our analysis to solely focus on predicting the interaction probability of the target behavior. Besides, in order to ensure the fairness of the comparison, for MMOE and PLE, we fix the number of experts in Tmall to 4, and 3 in Beibei and Taobao.
While MMOE and PLE exhibit a nearly uniform distribution of weights across all experts, PME selectively leverage information from different behaviors, thereby avoiding potential conflicts. 
This demonstrates the effectiveness of PME in utilizing diverse behavior information and improving overall performance.

\section{Conclusion}
\label{future works}

In this paper, we propose the Parallel Knowledge Enhancement based Framework (PKEF) for multi-behavior recommendation. To handle the problems of the existing multi-behavior approches, we devise Parallel Knowledge Fusion (PKF) module and Projection Disentangling Multi-Experts network (PME). PKF combines cascade and parallel paradigms to enhance behavior representations, addressing information bias caused by imbalanced behavioral interactions. PME treats each behavior as an independent task, generating specific expert information for each behavior using separate inputs. Besides, for each behavior, it leverages a projection mechanism to disentangle the shared and specific parts from other behaviors and aggregates the shared part while designing an auxiliary loss to further utilize the unique part. Thus, the negative transfer is significantly alleviated. Further, we perform extensive experiments on three real-world datasets to validate the effectiveness of our PKEF. The results provide further evidence of the rationale and effectiveness of the designed PKF and PME modules.

\begin{acks}

This work was partly supported by the Science and Technology Innovation 2030-Key Project (Grant No. 2021ZD0201404), Key Technology Projects in Shenzhen (Grant No. JSGG20220831110203007), Shenzhen Key Laboratory of next generation interactive media innovative technology (Grant No. ZDSYS20210623092001004) and Aminer·ShenZhen·ScientificSuperBrain. 

\end{acks}

%%
%% The acknowledgments section is defined using the "acks" environment
%% (and NOT an unnumbered section). This ensures the proper
%% identification of the section in the article metadata, and the
%% consistent spelling of the heading.
% \begin{acks}
% To Robert, for the bagels and explaining CMYK and color spaces.
% \end{acks}

%%
%% The next two lines define the bibliography style to be used, and
%% the bibliography file.
\bibliographystyle{ACM-Reference-Format}
\balance
\bibliography{sample-base}

%%
%% If your work has an appendix, this is the place to put it.
\appendix

\end{document}